\documentclass{appolb}
\usepackage{epsfig}
\begin{document}
% \eqsec  % uncomment this line to get equations numbered by (sec.num) 
\title{Fission-Fragment Mass Distribution and Particle Evaporation at low 
Energies
\thanks{This paper is devoted to Professor Adam Sobiczewski on the occasion 
of his 70th birthday}}

\author{ Ch.\ Schmitt$^1$, J.\ Bartel$^1$, K.\ Pomorski$^{1,2}$, 
A.\ Surowiec$^2$
\\
{\it $^1$IReS -- IN2P3/CNRS and Universit\'e Louis Pasteur, Strasbourg, France}
\\
{\it $^2$Katedra Fizyki Teoretycznej, Uniwersytet M. C. Sk\l odowskiej,
                               Lublin, Poland}}
%\date{}
% 
%
\maketitle   
\begin{abstract}
Fusion-fission dynamics is investigated with a special emphasis on fusion 
reactions at low energy for which shell effects and pairing correlations 
can play a crucial role leading in particular to multi-modal fission. To 
follow the dynamical evolution of an excited and rotating nucleus we solve  
a 2-dimensional Langevin equation taking explicitly light-particle 
evaporation into account. The confrontation theory-experiment is demonstrated 
to give interesting information on the model presented, its qualities as well 
as its shortcomings.
\end{abstract}  

\newcounter{myequation} 
\setcounter{myequation}{0}
\def\theequation{\arabic{myequation}}
%%%%%%%%%%%%%%%%%%%%%%%%%%%%%%%%%%%%%%%%%%%%%%%%%%%%%%%%%%%%%%%%%%%%%%%%%%%%%%%
%%%%%%%%%%%%%%%%%%%%%%%%%%%%%%%%%%%%%%%%%%%%%%%%%%%%%%%%%%%%%%%%%%%%%%%%%%%%%%%
%%%%%%%%%%%%%%%%%%%%%%%%%%%%%%%%%%%%%%%%%%%%%%%%%%%%%%%%%%%%%%%%%%%%%%%%%%%%%%%
%%%
\section{Introduction}

The description of the dynamical evolution of a compound nucleus along its 
way to fission, i.e. from its rather compact ground-state shape to its 
scission configuration, represents an intricate problem. 
Many ingredients enter into the description of such a process, starting from 
a sufficiently precise account of the formation of the compound system, to 
the determination of the multi-dimensional energy landscape, to the coupling 
between the collective dynamics and the intrinsic degrees of freedom of the 
nucleus, to the concept used to describe light-particle evaporation which 
can occur all along the fission path. 
As a general microscopic treatment is completely out of scope, different 
theoretical approaches based on a more or less classical picture 
\cite{GPW79}-\cite{Fro98} have been proposed.

We  have developed such a model describing  
the time evolution of a highly excited rotating nucleus and its subsequent 
decay through symme\-tric fission with pre-fission light-particle 
emission \cite{PBR96}. The aim of the present paper is   
to extend our theory to lower energy. Through a comparison with the 
available experimental data, in particular   
fission-fragment mass distributions and neutron pre-scission multiplicities,
we hope to get some valuable 
information on the behavior of
transport coefficients at low energy.
                                                                   \\[ -5.0ex]
%  
%%%%%%%%%%%%%%%%%%%%%%%%%%%%%%%%%%%%%%%%%%%%%%%%%%%%%%%%%%%%%%%%%%%%%%%%%%%%%%%
%%%%%%%%%%%%%%%%%%%%%%%%%%%%%%%%%%%%%%%%%%%%%%%%%%%%%%%%%%%%%%%%%%%%%%%%%%%%%%%
%%%%%%%%%%%%%%%%%%%%%%%%%%%%%%%%%%%%%%%%%%%%%%%%%%%%%%%%%%%%%%%%%%%%%%%%%%%%%%%
%
\section{Evolution of an excited rotating nucleus towards fission}
To study the time evolution of an excited rotating nucleus, the
system is assumed to follow a stochastic Langevin equation of motion \cite{PoStru90} of 
one or several 
collective variables that describe in an appropriate and sufficiently flexible 
way the deformation of the nucleus along its path to fission . 
                                                                   \\[ -5.0ex]
%%%%%%%%%%%%%%%%%%%%%%%%%%%%%%%%%%%%%%%%%%%%%%%%%%%%%%%%%%%%%%%%%%%%%%%%%%%%%%%
%%%%%%%%%%%%%%%%%%%%%%%%%%%%%%%%%%%%%%%%%%%%%%%%%%%%%%%%%%%%%%%%%%%%%%%%%%%%%%%
%
\subsection{Description of nuclear shapes}

To describe the large variety of deformed shapes that can appear in the 
fission process, the Trentalange--Koonin--Sierk (TKS) nuclear shape 
parametrization  \cite{TKS80} is used. In the case of an axially 
symmetric system the nuclear surface is given by  
\addtocounter{myequation}{1}
\begin{equation}
   \rho_s^2\,(z) = R_0^2 \, \sum_{\ell=0}^\Lambda \alpha_{\ell} \, 
                             P_{\ell}\,\left (\frac{z-{\bar z}}{z_0} \right ) 
        = R_0^2 \, \sum_{\ell=0}^\Lambda \alpha_{\ell} \, P_{\ell}\,(u) \; ,
                       \hspace{ 0.3cm}  z_0 = \frac{2 R_0}{3 \alpha_0} \; , 
	               \hspace{0.3cm} u = \frac{z-{\bar z}}{z_0}
\label{eqon1}\end{equation}
with $2 z_0$ the elongation of the shape in z direction, ${\bar z}$ its 
geometrical center and $R_0$ the radius of the corresponding spherical 
nucleus. The deformation parameters $\alpha_\ell$ define the  
shape.

This parametrization is strongly related to the well known 
Funny Hills $\{c,h, \alpha\}$ parametrization 
%\cite{BDJ72,Pau73} recalled below : 
\cite{BDJ72} recalled below : 
\addtocounter{myequation}{1}
\begin{equation}
   \rho_s^2\,(z) = c^2 \, R_0^2 \left\{ \begin{array}{ll}
   (1 - u^2) \, (A + \alpha \, u + B \, u^2) \; , \;            & B \geq 0  \\ 
   (1 - u^2) \, (A + \alpha \, u) \; {\rm exp}(B \, c^3 \, u^2) \; ,\; & B < 0 
\end{array}\right.
\label{eqon2}\end{equation}
with $z_0 = c \, R_0$ and 
where $A$ and $B$ are related to $c$ and $h$ through
$$
   A = \frac{1}{c^3} - \frac{1}{5} B \; , \;\;\;\;\;\;\;\; 
     B = 2 \, h + \frac{1}{2} (c-1) \;\; .                  $$

We have tested the convergence of these parametrizations for the des\-cription 
of symmetric fission-barrier heights and compared it to the results obtained 
using the expansion of the nuclear surface in spherical harmo\-nics. The 
agreement with experiment was better with the TKS parametrization using  
3 parameters $\alpha_2$, $\alpha_4$, $\alpha_6$ than with the later 
including deformation parameters up to $\beta_{14}$, thus showing the fast 
convergence of the TKS parametrization. 
%
%%%%%%%%%%%%%%%%%%%%%%%%%%%%%%%%%%%%%%%%%%%%%%%%%%%%%%%%%%%%%%%%%%%%%%%%%%%%%%%
%%%%%%%%%%%%%%%%%%%%%%%%%%%%%%%%%%%%%%%%%%%%%%%%%%%%%%%%%%%%%%%%%%%%%%%%%%%%%%%
%
\subsection{Fission dynamics and Langevin equation}
Fission dynamics is investigated through the resolution of the Langevin
equation which for the generalized coordinates $q_i$ is given by 
\addtocounter{myequation}{1}
\begin{eqnarray}
  && \!\!\!\!\!\!\!\! 
     {dq_i\over dt} = \sum_{j} \; [M^{-1}(\vec q)]_{i\, j} \; p_j  
                                                          \nonumber\\[ -1.0ex]
  &&                                                \label{eqn:rho}\\[ -1.0ex]
  && \!\!\!\!\!\!\!\! 
     {dp_i\over dt} =  - {1\over 2} \sum_{j,k} \,
                             {d[M^{-1}(\vec q)]_{jk} \over dq_i} \; p_j \; p_k 
                    - {dV(\vec q)\over dq_i} 
              - \sum_{j,k} \gamma_{ij}(\vec q) \; [M^{-1}(\vec q)]_{jk} \; p_k 
%              - \sum_{j,k}  [\gamma(\vec q)]_{ij} \; [M^{-1}(\vec q)]_{jk} \; p_k 
                    + F_i(t)                                     \nonumber
\end{eqnarray}
${}$
                                                                   \\[ -3.0ex]
where $p_i$ are the canonical momenta associated with the coordinates $q_i$. 
$[M(\vec q)]$ represents the tensor of inertia determined in our approach in 
the irrotational incompressible fluid approximation of Werner-Wheeler as 
deve\-loped by Davies, Sierk and Nix \cite{Da76} and $[\gamma(\vec q)]$ 
corresponds to the friction tensor calculated in the framework of the so-called 
{\it wall and window friction} model \cite{BRS77,Fe87}. 
The collective potential $V(\vec q)$ is defined in our approach as the 
Helmholtz free energy at given deformation \cite{PBR96,BMR96}. The term 
$F_i(t)$ stands for the random Langevin force 
which couples collective dynamics to the intrinsic degrees of freedom. We have 
$F_i(t) \!\!=\!\! \sum_{j} g_{ij}(\vec q) \; G_j(t)$ where the strength 
tensor $[g(\vec q)]$ is given by the diffusion tensor  $[D(\vec q)]$ through 
$D_{ij} \!\!=\!\! \sum_{k} g_{ik} \; g_{jk}$ and $\vec G(t)$ is a stochastic 
function. 
In our model it is assumed that diffusion is related to friction through the 
Einstein relation $[D(\vec q)] \!=\! [\gamma(\vec q)] \; T$ where T corresponds 
to the nuclear temperature \cite{PBR96}. The explicit expressions of these 
quantities in the TKS parametrization have been presented and discussed in 
details in \cite{BMR96}.

The friction model we are using is based on a classical concept  valid 
 at high energy. When going to lower temperatures   
this picture can only be considered as an upper limit since  
  nucleon-nucleon collisions 
become less and less frequent thus reducing friction \cite{PH91}.  
We also know that the Einstein relation is in principle only valid at high 
energy \cite{HoIv99}. We shall come back to 
 these approximations in section 4.5.\ and show that one has  
to modify this simplified description at low temperature  
to correctly describe the experimental data.

Another quantity entering the Langevin equation and whose tempe\-ra\-ture 
dependence 
requires special attention is the potential $V(\vec q\,)$ namely because of the vanishing 
of quantal effects at high excitation energy. In our approach valid up to now 
for symmetric fission, it consisted of a 
temperature dependent Liquid Drop Model (LDM) term only. At lower energy  we have to add to 
this macroscopic contribution the shell corrections which are evaluated 
at each deformation using the Strutinsky's approach \cite{S67} and the   
pairing correlations which we calculate in the framework of the BCS model \cite{BCS57} with 
a constant pairing strength ({\it seniority scheme}) \cite{NTS69}. 

The generalized coordinates $q_i$ which enter the Langevin equation are either 
chosen as the deformation parameters generating the nuclear shape (e.g. 
coefficients $\alpha_{\ell}$) or as more physically relevant  
quantities (elongation, mass asymmetry, etc) 
which are determined through these parameters \cite{BMR96}.

Up to now \cite{PBR96,PNS00} we have investigated the case of highly excited 
compound nuclei giving rise to symmetric fission. Such a process can be 
des\-cribed approximately by a single collective coordinate characterizing the 
nuclear elongation as explained in ref.\ \cite{BMR96}. This approach has been 
proven quite successful reproducing experimental pre-scission neutron  
multiplicities with an accuracy of $10 - 20 \%$ for nuclei ranging from 
$^{126}$Ba to the region of superheavy elements \cite{PNS00}. As our aim in the present 
paper is to investigate systems at  
  lower energy, one has to be able to describe {\it multi-modal} fission caused by the 
  competition between symmetric and asymmetric splitting generated by the  
  quantal effects present at low temperature. Dealing with asymmetric 
  shapes, we need to take at least two 
collective variables (e.g. elongation and asymmetry) into account 
describing the compound nucleus along its 
deformation process. For this purpose we choose to use the Funny-Hills
parametrization and to restrict ourselves to the 2-dimensional ($c, \alpha$) 
deformation space imposing 
$h = 0$.  Indeed, one can show that the influence of
the {\it neck parameter} $h$ can be considered as rather small, 
at least in the semi-classical limit \cite{CSLi00,CS02}.
                                                                   \\[ -5.0ex]
%
%%%%%%%%%%%%%%%%%%%%%%%%%%%%%%%%%%%%%%%%%%%%%%%%%%%%%%%%%%%%%%%%%%%%%%%%%%%%%%%
%%%%%%%%%%%%%%%%%%%%%%%%%%%%%%%%%%%%%%%%%%%%%%%%%%%%%%%%%%%%%%%%%%%%%%%%%%%%%%%
%
\subsection{Entrance channel effects}
In order to solve the Langevin equation of motion one needs to specify the 
initial conditions of the trajectory (for reasonable statistics we need to 
consider $10^4$ to $10^6$ trajectories) from which the compound system starts 
and evolves either through the fission channel or ending up as an evaporation 
residue. The initial conditions for  $\vec q_0$ and $\vec p_0$ are fixed to 
the ground-state deformation 
and drawn from a normalized gaussian distribution respectively \cite{PBR96}. The nuclear 
systems we have investigated so far   
were generated through heavy-ion collisions which can lead to a large variety of the angular momentum of the
synthesized nucleus. The  initial spin distribution of the former is determined in our model by 
solving a Langevin equation \cite{Prz94} describing the evolution of the two colliding ions from an
infinite distance up to fusion. 
The Langevin equation (3) is then solved in order to describe the dynamical 
evolution of the synthesized nucleus taking particle emission into account by 
coupling the Langevin equation to the Master equations governing this 
evaporation process. 
For each trajectory we start with a given compound system characterized by its 
excitation energy and angular momentum. The final prediction, which can be 
compared to experiment, is then determined by weighting the calculations made 
at given angular momentum by the fusion-fission cross section \cite{PNS00}.
${}$
                                                                   \\[ -18.0ex]
\begin{center}
\vspace{-1.0cm}
\hspace{0.0cm}\epsfig{file=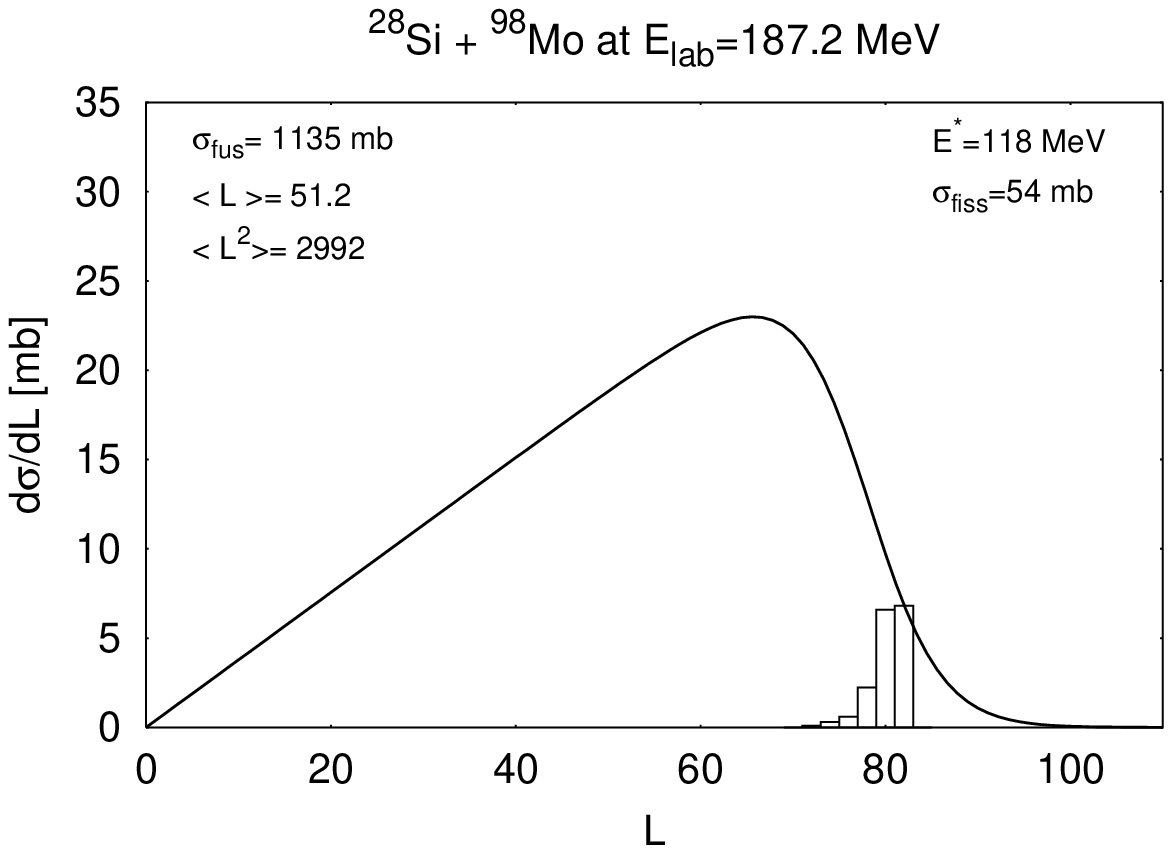,width=8.0cm,height=6.5cm}
\end{center}
\vspace{-0.3cm}
FIG.\ 1. Differential fusion (solid line) and fission (histogram) 
cross-section for the reaction $^{28}$Si$ + ^{98}$Mo$ \rightarrow ^{126}$Ba at 
$E^*_{tot} = 118.5$ MeV.
\vspace{0.1cm}
\begin{center}
%\hspace*{-0.05cm}\epsfig{file=bar-height-L.eps,height=6.0cm,width=6.7cm}
\hspace*{-0.05cm}\epsfig{file=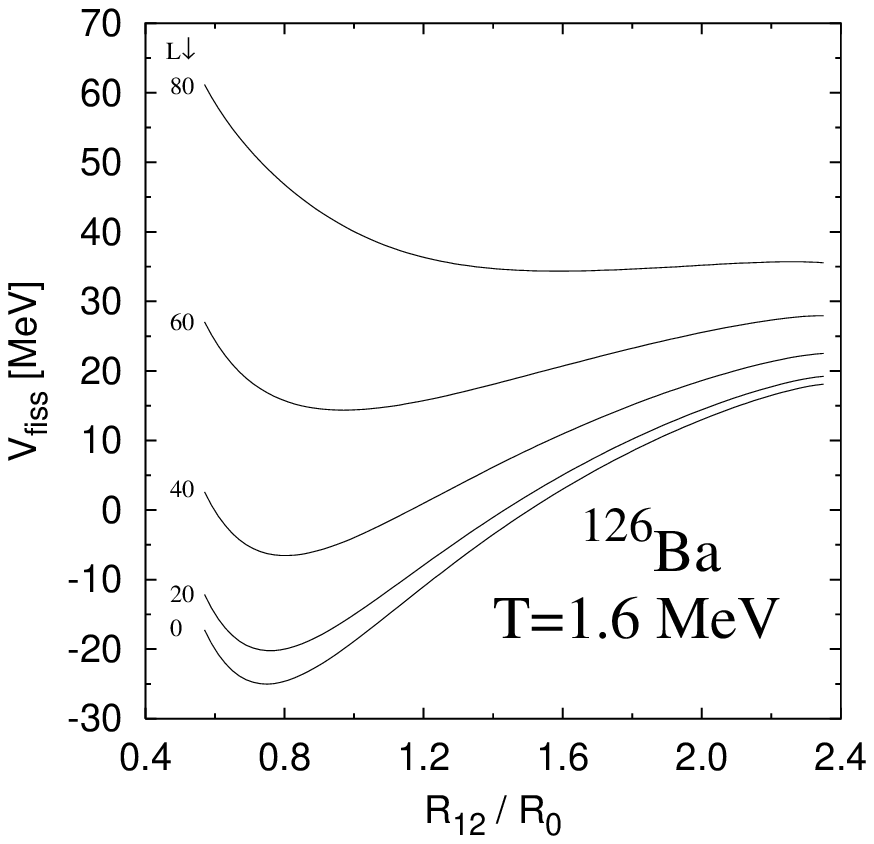,height=6.0cm,width=6.7cm}
\vspace*{-6.0cm}
%\hspace*{-1.50cm}\epsfig{file=bar-height-T.eps,height=6.0cm,width=6.7cm}
\hspace*{-1.50cm}\epsfig{file=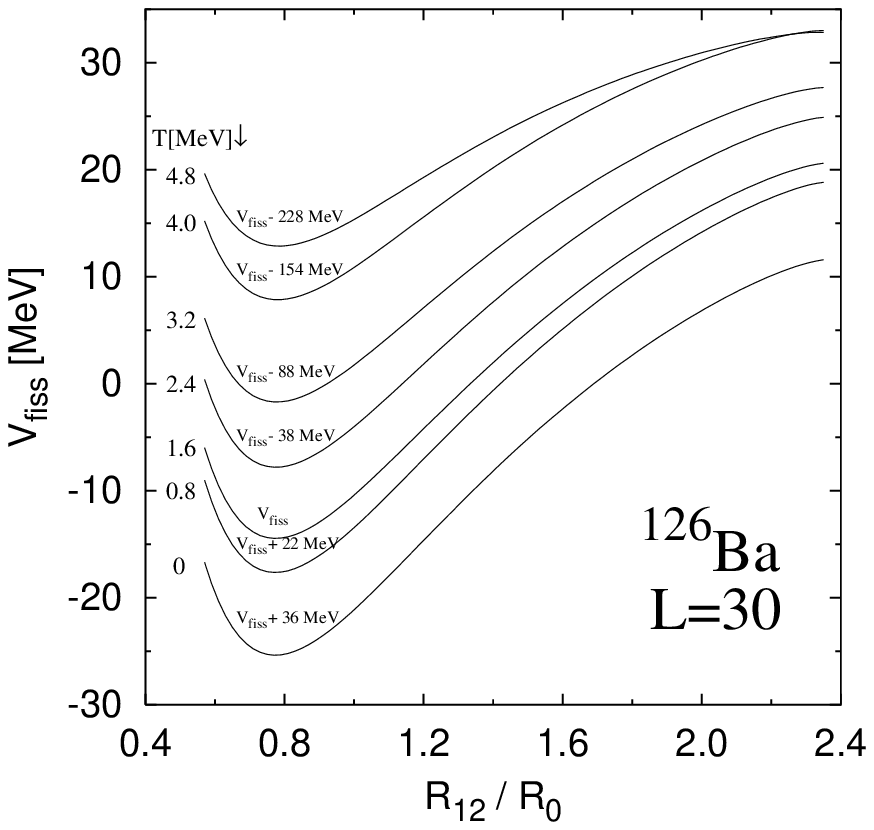,height=6.0cm,width=6.7cm}
\end{center}
\vspace{5.6cm}
FIG.\ 2. Fission barriers for the nucleus $^{126}$Ba as function 
of angular momentum at fixed total excitation energy (left) and of temperature at fixed 
angular momentum (right).
                                                                   \\[ -1.0ex]

Fig.\ 1 shows the fusion and fission cross-sections obtained for the system  
$^{28}$Si$ + ^{98}$Mo$ \rightarrow ^{126}$Ba at a total excitation energy of $E^*_{tot} = 118.5$ MeV. 
One notices that fission yields are rather 
small and located in the tail of the spin distribution at high values of the 
angular momentum where fission barriers are low. A study of the 
fission-barrier height as function of angular momentum and thermal excitation 
energy is given in Fig.\ 2 from which we conclude  
 that a careful description of the fusion 
cross-section through its initial spin distribution is necessary if one wants to   
 describe the competition between the decay by fission and light-particle evaporation.
                                                                   \\[ -4.0ex]
% 
%%%%%%%%%%%%%%%%%%%%%%%%%%%%%%%%%%%%%%%%%%%%%%%%%%%%%%%%%%%%%%%%%%%%%%%%%%%%%%%
%%%%%%%%%%%%%%%%%%%%%%%%%%%%%%%%%%%%%%%%%%%%%%%%%%%%%%%%%%%%%%%%%%%%%%%%%%%%%%%
%%%%%%%%%%%%%%%%%%%%%%%%%%%%%%%%%%%%%%%%%%%%%%%%%%%%%%%%%%%%%%%%%%%%%%%%%%%%%%%
%
\section{Light-particle emission}
${}$
                                                                   \\[ -5.0ex]
								   
Fission dynamics of an excited rotating nucleus usually goes along with the 
emission of light particles (we will consider neutrons, protons and $\alpha$ 
particles). This evaporation process is governed by the emission width 
$\Gamma_{\nu}^{\mu \kappa}(E^*,L)$ for emitting a particle of type $\nu$, 
energy $\varepsilon_{\mu}$ and angular momentum $\ell_{\kappa}$ from a  
 nucleus characterized by its thermal excitation energy $E^*$ and its rotational angular 
momentum $L$. In order to determine $\Gamma_{\nu}^{\mu \kappa}(E^*,L)$ we use 
two different prescriptions.
\begin{center}
\hspace{0.0cm}\epsfig{file=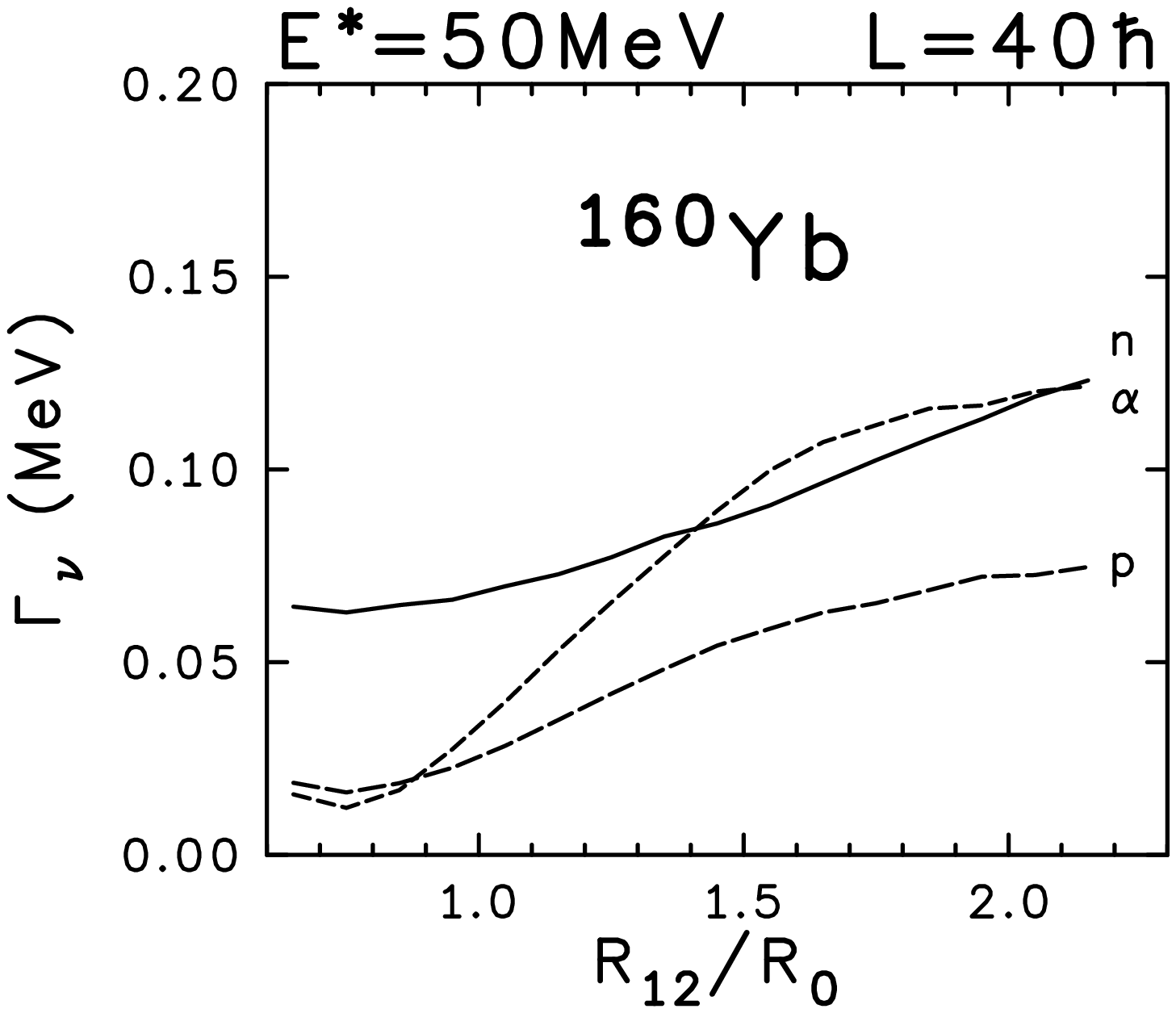,width=9.0cm,height=5.0cm}
\end{center}
\vspace{-0.3cm}
FIG.\ 3. Emission widths for neutrons, protons and $\alpha$ particles emitted 
         from the system $^{160}$Yb (E$^*$=50 MeV, L=40$\hbar$) as a function 
	 of elongation.
\vspace{ 0.3cm}

In Weisskopf's evaporation theory 
\cite{Wei37}  the decay rates are essentially evaluated through 
the level densities of the mother and the daughter 
nuclei and the transmission coefficient for emitting the particle 
from a given point of the nuclear surface into a given direction as explained 
in ref.\ \cite{PBR96}. In
 practice it is not possible to discuss the values of the emission width
  for each energy, angular
momentum and  position of the emission point on the nuclear surface. We therefore
use them to determine the probability $\Gamma_{\nu} (E^*,L)$ of emitting a given particle  from
a given nucleus at given deformation. This simplified procedure calculates a 
transmission coefficient obtained by an averaging over the different 
emission directions and over the whole surface of the deformed  
nucleus. A detailed description of 
this procedure can be found in ref.\ \cite{PBR96}. Also other groups 
\cite{SDP91,AAR96,Fro98}
have dealed with particle emission 
in connection with fission dynamics but, to our knowledge,  
none of them has taken nuclear deformation explicitly into account 
as we have, even if it is in an approximate way.
In Fig.\ 3 the evaporation rate $\Gamma_{\nu}$ is displayed 
for a hot rotating nucleus $^{160}$Yb. It becomes obvious 
that the deformation 
dependence of $\Gamma_{\nu}$ is essential 
and that assuming a deformation independent emission width could probably lead   
to wrong predictions.

The second approach we used so far to describe particle emission 
calculates the transition rates $\Gamma^{\mu\kappa}_\nu$ 
through the probability that a particle which impinges on the nuclear 
surface at a given point $\vec r_0\,'$ and with a given velocity $\vec v_0\,'$ is 
actually transmitted \cite{DPR95}. 
In this framework the quantity $\Gamma^{\mu\kappa}_\nu$ is determined as 
\begin{equation}
  \Gamma^{\mu\kappa}_\nu = {d^2 n_\nu \over d\varepsilon_\mu d\ell_\kappa}  
                             \Delta\varepsilon \,\Delta \ell \; .
\label{eqtw10}\end{equation}

The number $n_\nu$ of particles of type $\nu$ which are emitted per time
unit through the surface $S$ of the fissioning nucleus is given by
$$
n_\nu = \int\limits_S \! d\sigma \int \! d^3 p' \, f_\nu(\vec r_0\,', 
 \vec p\,') \, v'_\perp(\vec r_0\,') \, w_\nu( v'_\perp (\vec r_0\,'))
 $$ 
where the quantity $f_\nu(\vec r\,' , \vec p\,')$ corresponds to the 
quasi-classical phase-space distribution function \cite{DPR95}.
                                                                   \\[  0.0ex]
\vspace*{1.0cm}
\hspace*{0.0cm}\epsfig{file=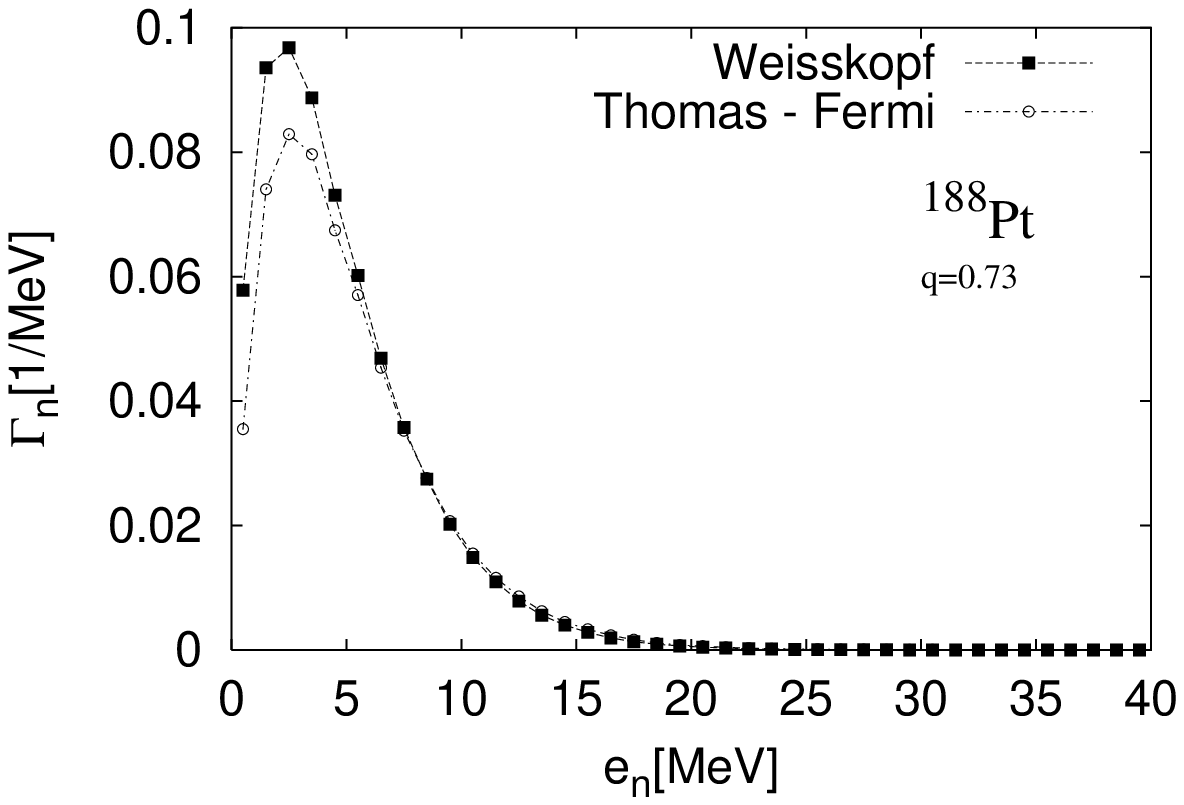,height=5.3cm,width=6.0cm,angle=0}  
\vspace*{-6.0cm}
\hspace*{0.0cm}\epsfig{file=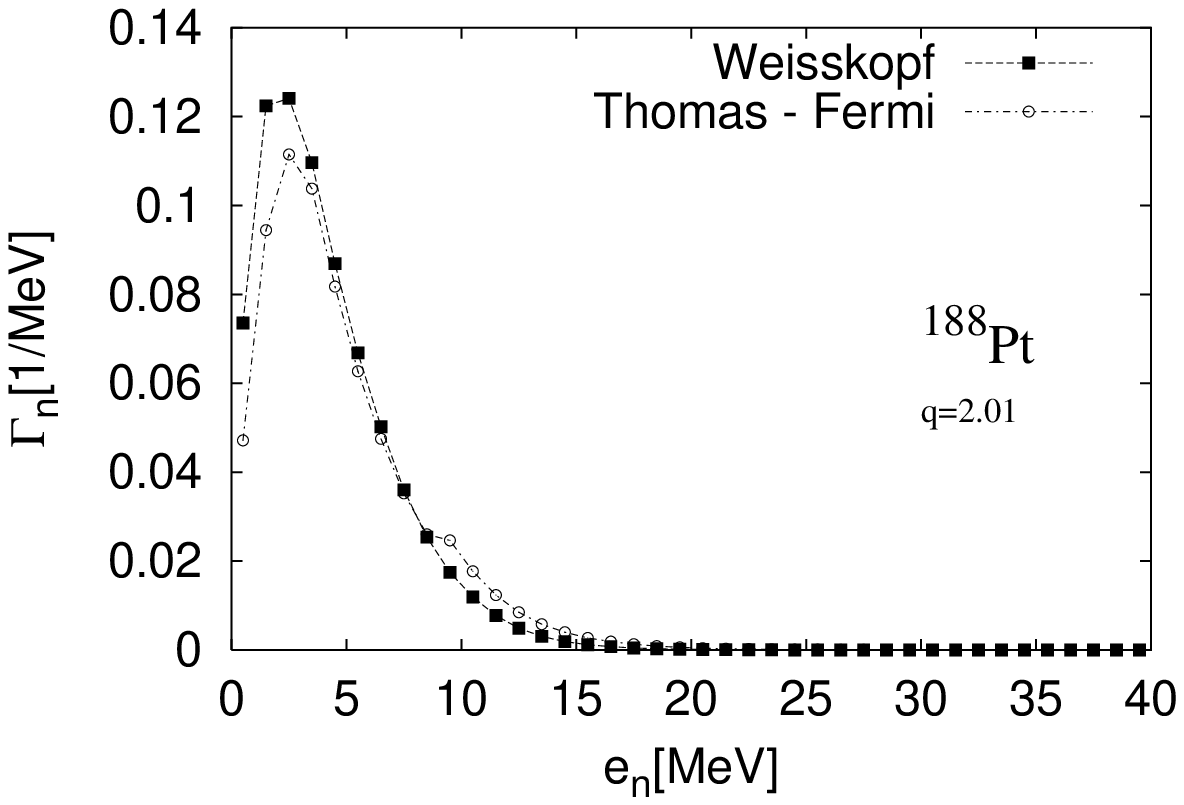,height=5.3cm,width=6.0cm,angle=0}  
%\hspace*{0.0cm}\epsfig{file=../totalite/wefgefn073.eps,height=5.3cm,width=6.0cm,angle=0}  
%\vspace*{-6.0cm}
%\hspace*{0.0cm}\epsfig{file=../totalite/wefgefn201.eps,height=5.3cm,width=6.0cm,angle=0}  
%
${}$
                                                                   \\[ 27.0ex]
\vspace*{1.0cm}
%\hspace*{0.0cm}\epsfig{file=../totalite/wefgefp073.eps,height=5.3cm,width=6.0cm,angle=0}  
%\vspace*{-6.0cm}
%\hspace*{0.0cm}\epsfig{file=../totalite/wefgefp201.eps,height=5.3cm,width=6.0cm,angle=0}  
\hspace*{0.0cm}\epsfig{file=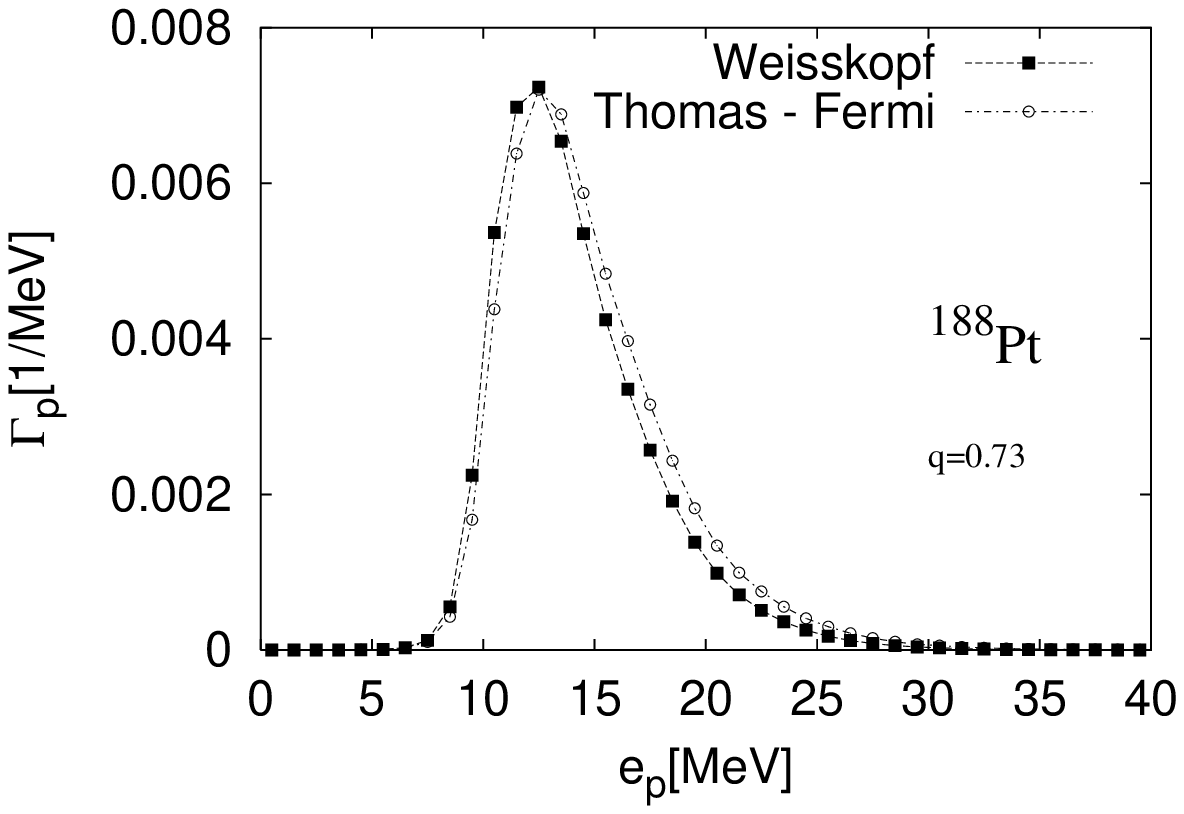,height=5.3cm,width=6.0cm,angle=0}  
\vspace*{-6.0cm}
\hspace*{0.0cm}\epsfig{file=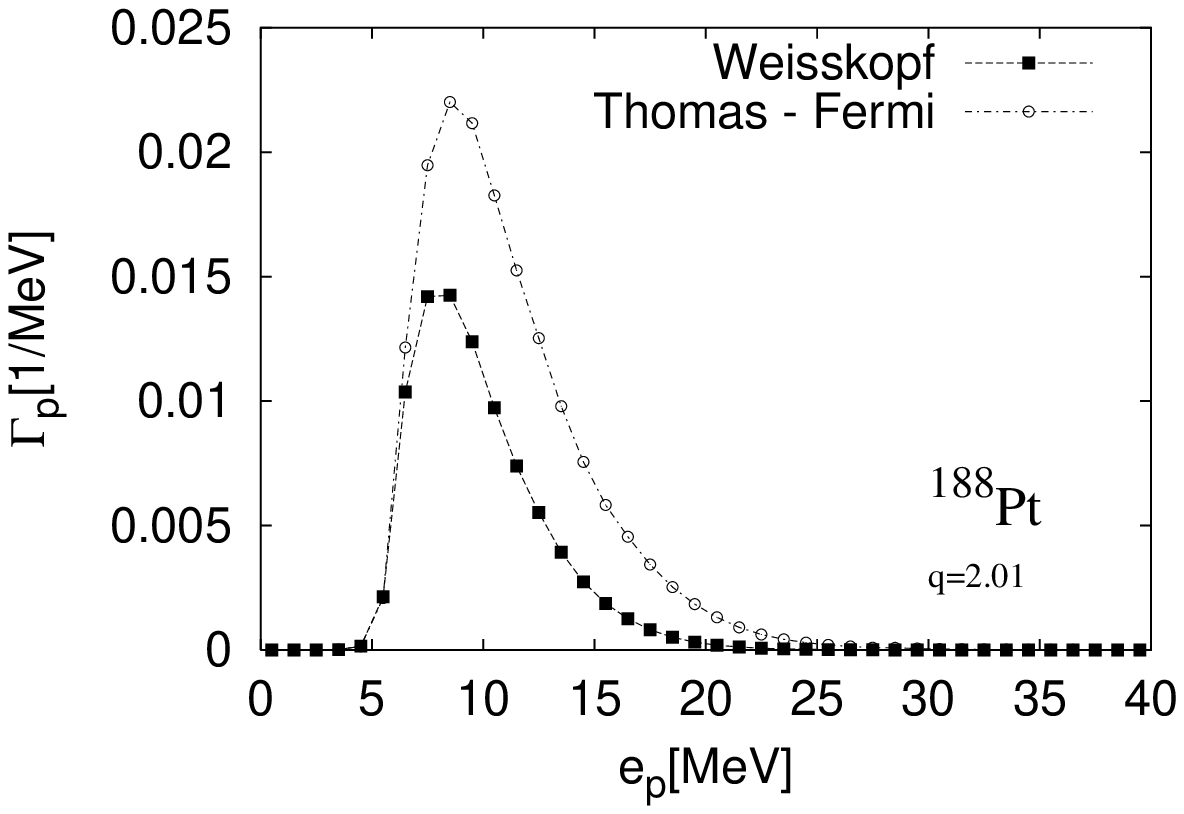,height=5.3cm,width=6.0cm,angle=0}  
${}$
                                                                   \\[ 30.0ex]
FIG.\ 4. Emission rates $\Gamma_n$ and $\Gamma_p$ 
for neutrons and protons obtained in the Weisskopf and the 
distribution function (called Thomas-Fermi here) approaches at deformations close to the spherical 
shape ($q=0.73$) and to the scission configuration ($q=2.01$).
\vspace{ 0.2cm}

Fig.\ 4 gives a comparison of the neutron and proton emission rates in the 
two evaporation models obtained for different values of a collective 
coor\-dinate $q$ related to nuclear elongation \cite{BMR96} for the system 
$^{188}$Pt 
($E^* \!\!=\!\! 100$ MeV, $L \!\!=\!\! 0 \hbar$) \cite{SPS01}. Both models 
yield emission rates that are reasonably close for both types of particles 
for all elongations except for an increase of the distribution function 
approach relative to the Weisskopf prediction in the case of protons for very 
large deformations. 
One should also notice the deformation dependence of the proton 
emission width $\Gamma_p$ that can be easily understood if 
one keeps in mind that the Coulomb barrier which charged particles have 
to overcome depends on the direction of the emission  
(an emission along the tips is favored compared to an emission 
perpendicular to the symmetry axis).

The determination of the phase-space distribution function is quite intricate 
in the case of $\alpha$-particles which are composite particles. 
We are presently working on a model which determines the $\alpha$-particle 
distribution function $f_\alpha$ through those of two correlated protons 
and neutrons respectively \cite{SPS01}.
                                                                   \\[ -4.0ex]
%%%%%%%%%%%%%%%%%%%%%%%%%%%%%%%%%%%%%%%%%%%%%%%%%%%%%%%%%%%%%%%%%%%%%%%%%%%%%%%
%%%%%%%%%%%%%%%%%%%%%%%%%%%%%%%%%%%%%%%%%%%%%%%%%%%%%%%%%%%%%%%%%%%%%%%%%%%%%%%
%%%%%%%%%%%%%%%%%%%%%%%%%%%%%%%%%%%%%%%%%%%%%%%%%%%%%%%%%%%%%%%%%%%%%%%%%%%%%%%
%
\section{Theoretical results of fission dynamics}
\subsection{Influence of quantal effects}
\subsubsection{One- versus 2-dimensional Langevin equation}

In the framework of the 2-dimensional Langevin equation solved in the 
($c, \alpha$) deformation space, the LDM energy landscape is displayed on 
Fig.\ 5 together with a typical fission trajectory for the compound nucleus 
$^{227}$Pa at a total excitation energy of $E^*_{tot} \!=\! 26$ MeV and an 
angular momentum of $L = 60 \hbar$. We choose this specific nuclear system 
because it was the object of a recent experimental campaign \cite{CS02}.
As no shell effects are taken into account here, only   
the symmetric fission valley is present. Consequently the compound nucleus 
starting from its ground-state  
deformation ($c \!=\! 1.11 , \alpha \!=\! 0$), naturally ends up in
the symmetric fission channel. In this calculation we have not coupled particle 
emission to the Langevin equation and therefore cannot make any statement on 
particle multiplicities. The 
fission time, defined as the average time which a trajectory takes 
to reach the scission point, is in the present 2-dimensional 
treatment reduced by about 7\% (5.96 $10^{-17}$ sec versus 6.36 $10^{-17}$ sec) as 
compared to its 1-dimensional 
value \cite{CSLi00}.

Let us try to understand this result since it might seem astonishing 
that resolving the 2-dimensional Langevin equation, where trajectories can 
fill out more effectively the deformation space (as it is demonstrated with the typical trajectory drawn on
Fig.\ 5), would lead to shorter fission 
times than when the compound nucleus follows the 
deepest symmetric fission valley of the 1-dimensional picture. In fact we have to think of the Langevin 
equation as an approximation to the Fokker-Planck one which deals with probability 
distributions. In an 1-dimensional space the system is constrained whereas, the more 
 the dimensionality is increased, the less constraints one has.
                                                                   \\[ -6.0ex]
\begin{center}
\epsfig{file=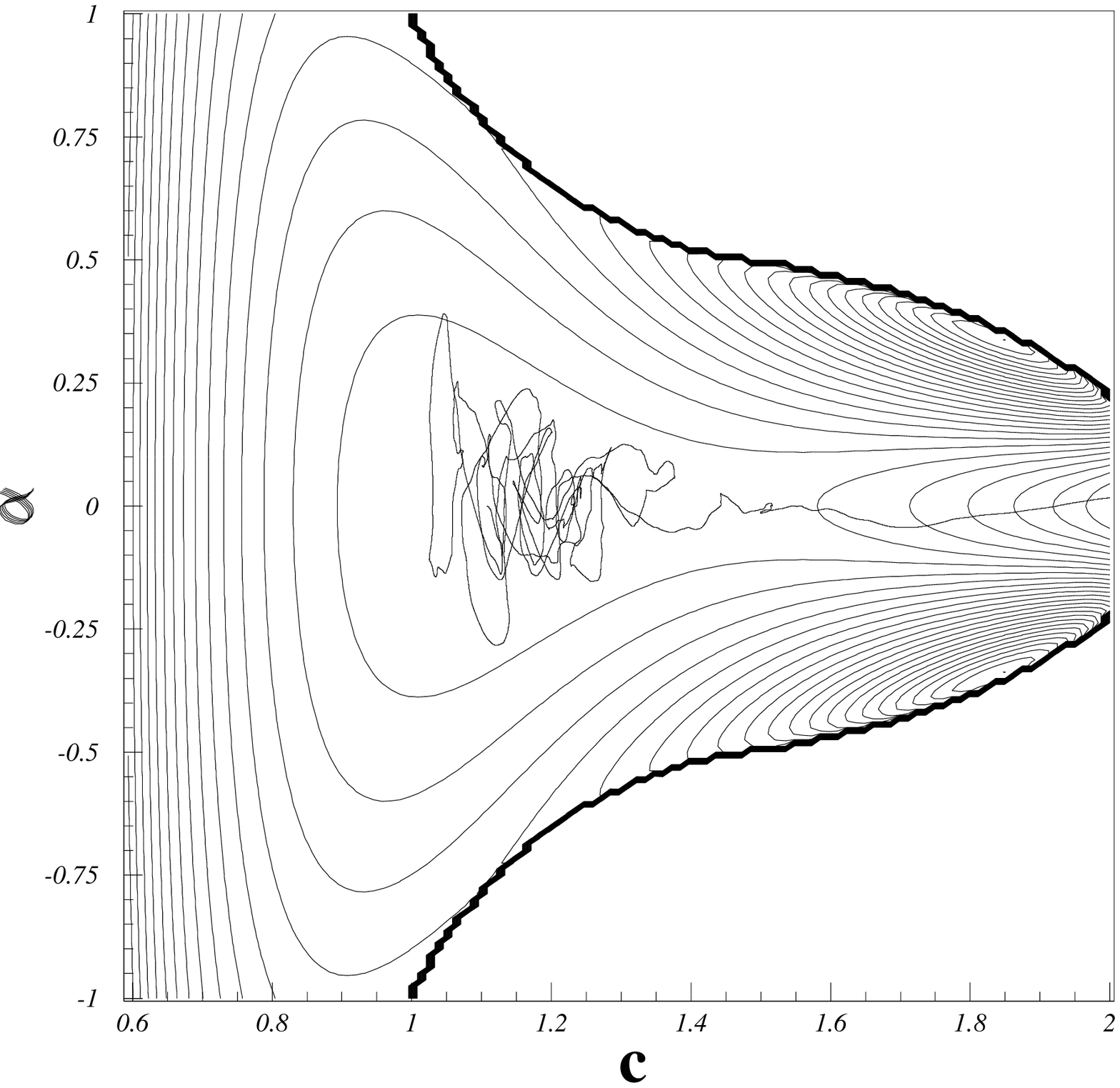,width=9.0cm,height=6.50cm}
\end{center}
\vspace{-0.4cm}

FIG.\ 5. Temperature dependent LDM energy landscape and typical fission 
trajectory for the compound system $^{227}$Pa 
($E^*_{tot} = 26$ MeV, $L = 60 \hbar$). 
\vspace{-0.1cm}
  
This result on fission times seems also to indicate that in the case of 
highly excited nuclei our previous 
1-dimensional description was already fairly accurate. The small change 
in times should indeed imply a rather small change in pre-fission 
particle multiplicities.  
\subsubsection{Influence of shell effects and pairing correlations}

Let us now go one step further by including in our potential energy calculation
quantal effects and their dependence on temperature. It is ge\-ne\-rally  
admitted that shell corrections have disappeared for temperatures above 
$2.5$ to $3$ MeV whereas pairing correlations have already vanished at 
$T \!\approx\! 1.5$ MeV or even before. In order to take care of the  
$T$-dependence of quantal corrections we multiply their values obtained at 
$T \!=\! 0$ MeV with a temperature smoothing function which goes to zero at 
$T \!\!=\! 3$ MeV for shell corrections and at $T \!=\! 1.5$ MeV in the case of 
pairing \cite{I78} - \cite{EgRo00}. The energy landscape then obtained 
for $E^*_{tot} \!=\! 26\,$MeV and $L \!=\! 60\,\hbar$ is drawn on Fig.\ 6. 
Comparing the landscapes in Figs.\ 5 and 6, one notices the appearance, due 
to the presence of microscopic corrections, of asymmetric fission channels 
beyond $c \approx 1.7$ ending up in well pronounced valleys around 
$\alpha \approx \pm 0.035$. 
								   
The resolution of the Langevin equation in the landscapes of Figs.\ 5 and 6 
gives rise to the distributions for the asymmetry parameter and 
fission-fragment masses presented in Figs.\ 7 and 8 respectively. Whereas 
symmetric distributions were obviously expected for the pure LDM landscape, 
the distributions obtained in the case where quantal effects are present are 
a little surprising, because of their very strong asymmetry in spite of the 
rather flat energy landscape of Fig.\ 6 in the asymmetry direction $\alpha$ 
for large elongations ($c \approx 2.0$). However one should not forget that 
the fragment mass distribution is decided all along the fission path and not 
only in the immediate neighborhood of the exit point \cite{MN70} - 
\cite{NPPW79}. As the asymmetric valley is around $1$ MeV deeper than the 
symmetric fission path in the vicinity of $c \approx 1.8 - 1.9$ where 
$\alpha \approx \pm 0.035$, the predominant part of the trajectories finally ends up 
in this asymmetric channel.
                                                                   \\[ -4.5ex]
%\vspace{-0.8cm}
\begin{center}
%\hspace{-14.6cm}\epsfig{file=landsc-2dim.ps,height=1.40cm,width=1.40cm,angle=270}
%\hspace{-10.6cm}\epsfig{file=landsc-2dim.ps,height=1.0cm,width=1.0cm,angle=270}
\hspace{-1.0cm}\epsfig{file=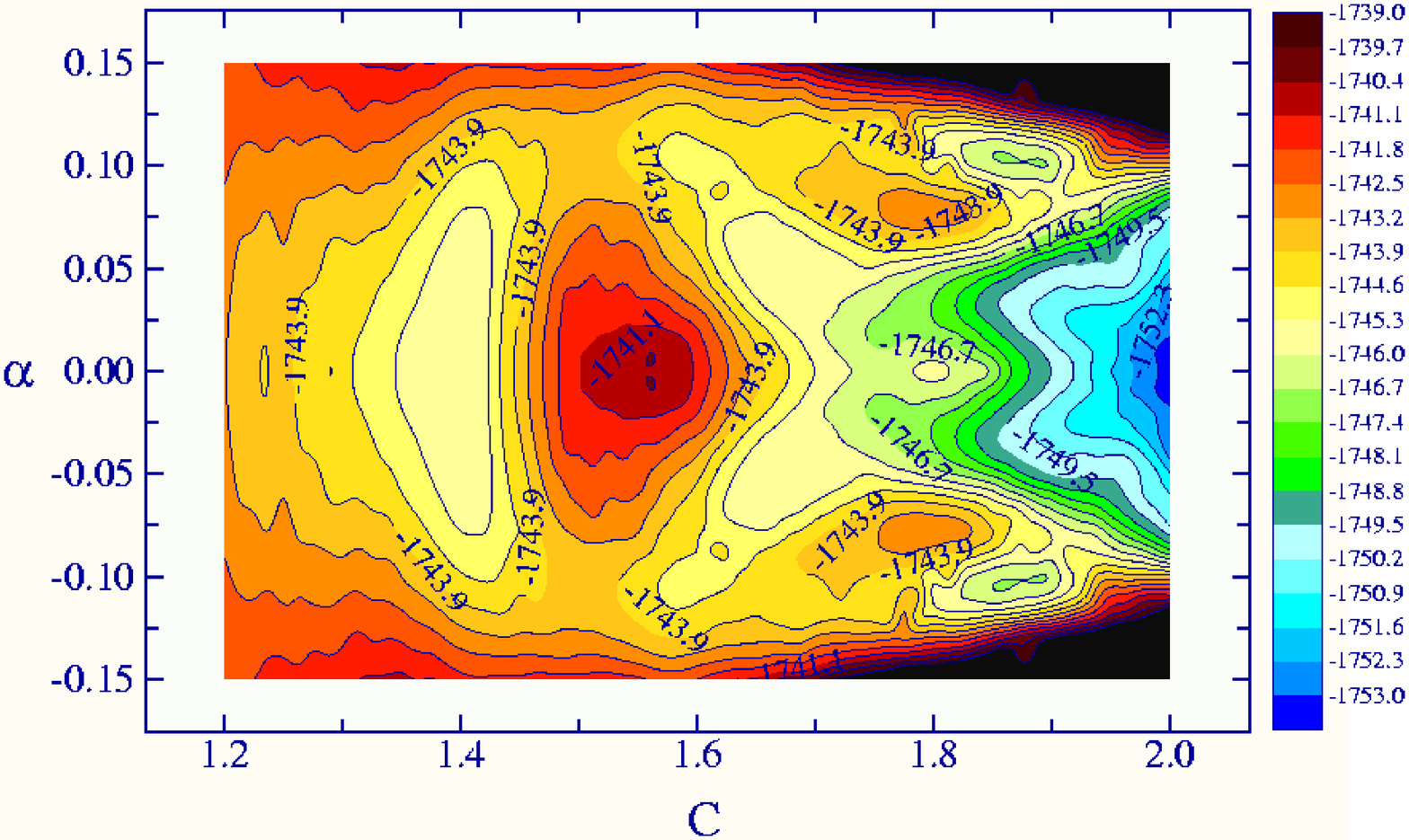,width=12.0cm,angle=0}
\end{center}
\vspace{-0.70cm}
FIG.\ 6. Same as Fig.\ 5 but with inclusion of quantal effects. 
                                                                   \\[ -1.0ex]
\vspace{-0.5cm}
\begin{center}
\epsfig{file=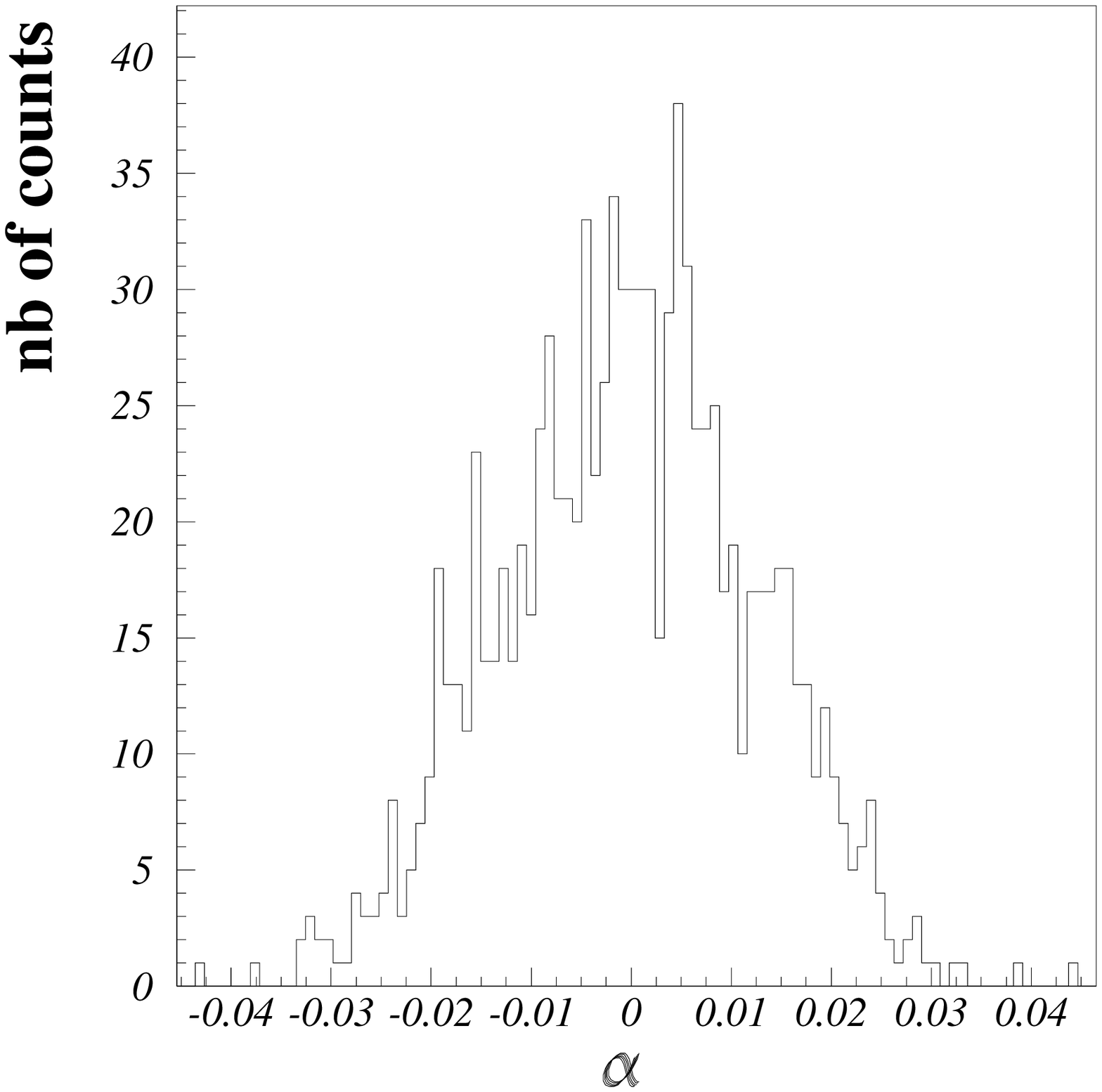,height=5.5cm,width=6.0cm}
\epsfig{file=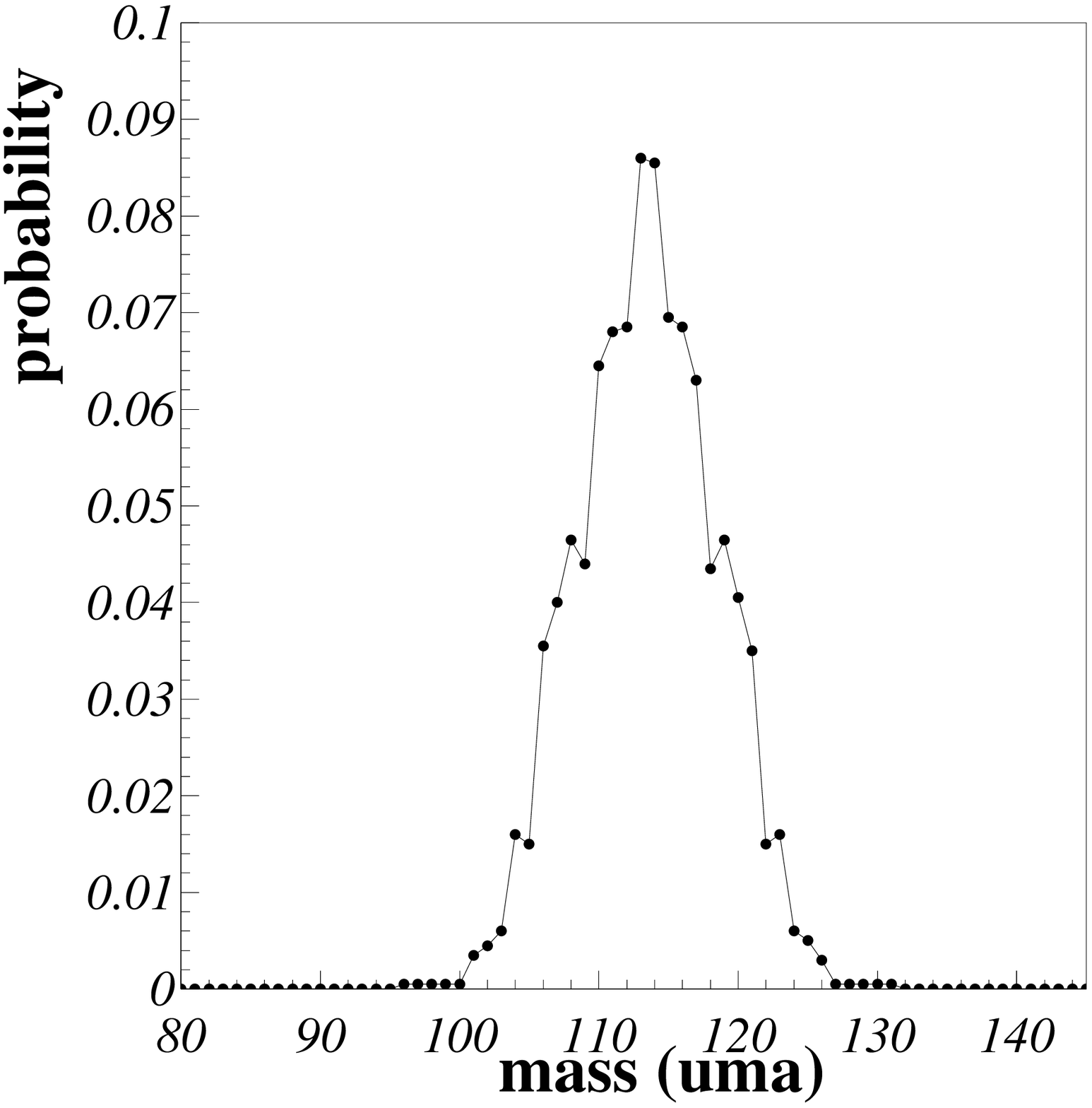,height=5.5cm,width=6.0cm}
\end{center}
\vspace{-0.3cm}
FIG.\ 7. Fission-fragment distribution as function of the asymmetry parameter 
$\alpha$ (left) and of the fragment mass (right) when quantal effects are 
omitted. 
                                                                   \\[ -3.0ex]
\begin{center}
\epsfig{file=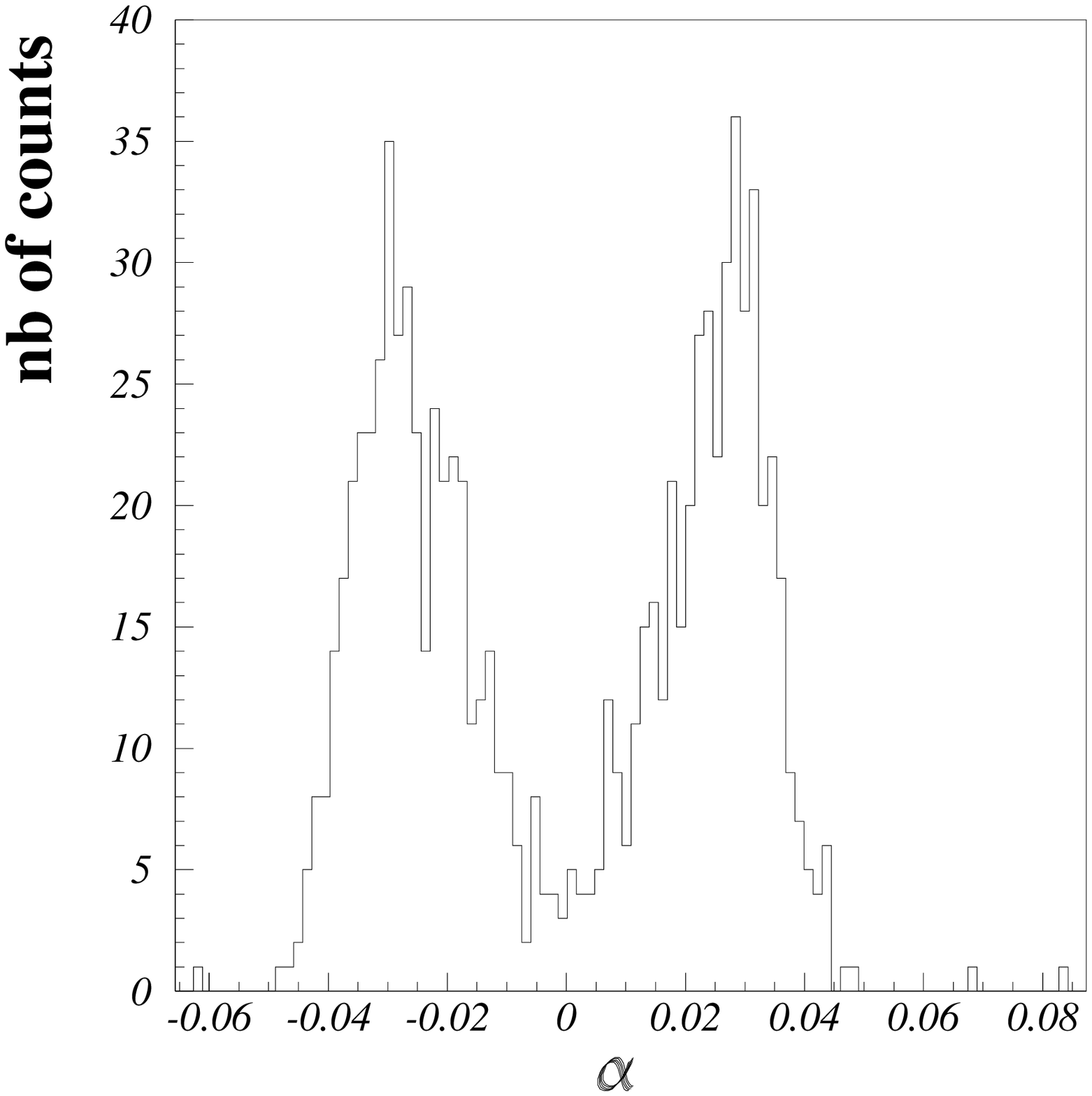,height=6.00cm,width=6.0cm}
\epsfig{file=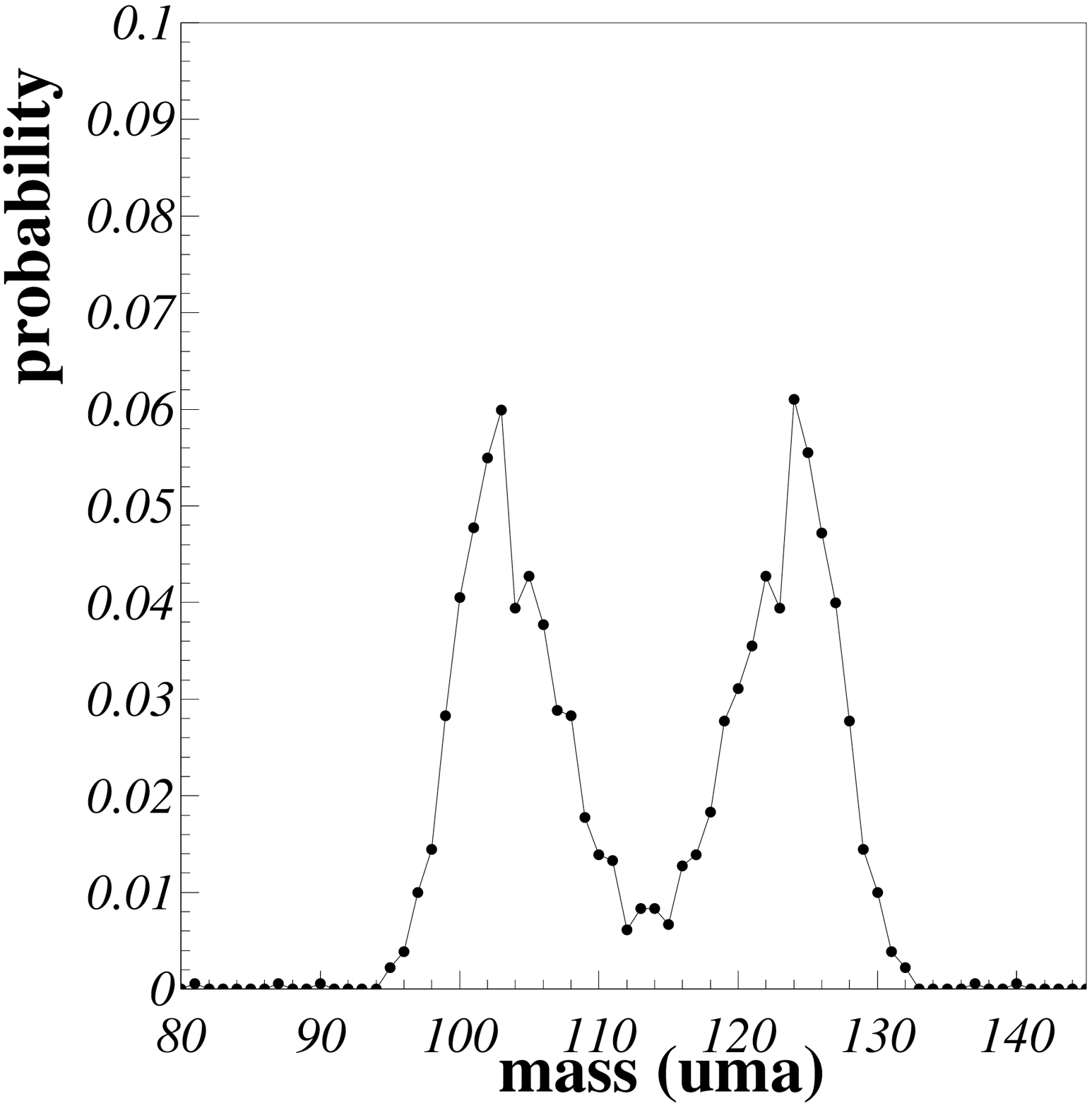,height=6.00cm,width=6.0cm}
\end{center}
\vspace{-0.3cm}
FIG.\ 8. Same as Fig.\ 7 but with inclusion of quantal effects.
                                                                   \\[ -1.0ex]

It is interesting to notice that the average fission time 
is increased from 8.0 $10^{-17}$ sec to 16.2 $10^{-17}$ sec when going from 
the LDM picture to the one with shell and pairing corrections. 
                                                                    \\[ -5.0ex]
\begin{center}
\epsfig{file=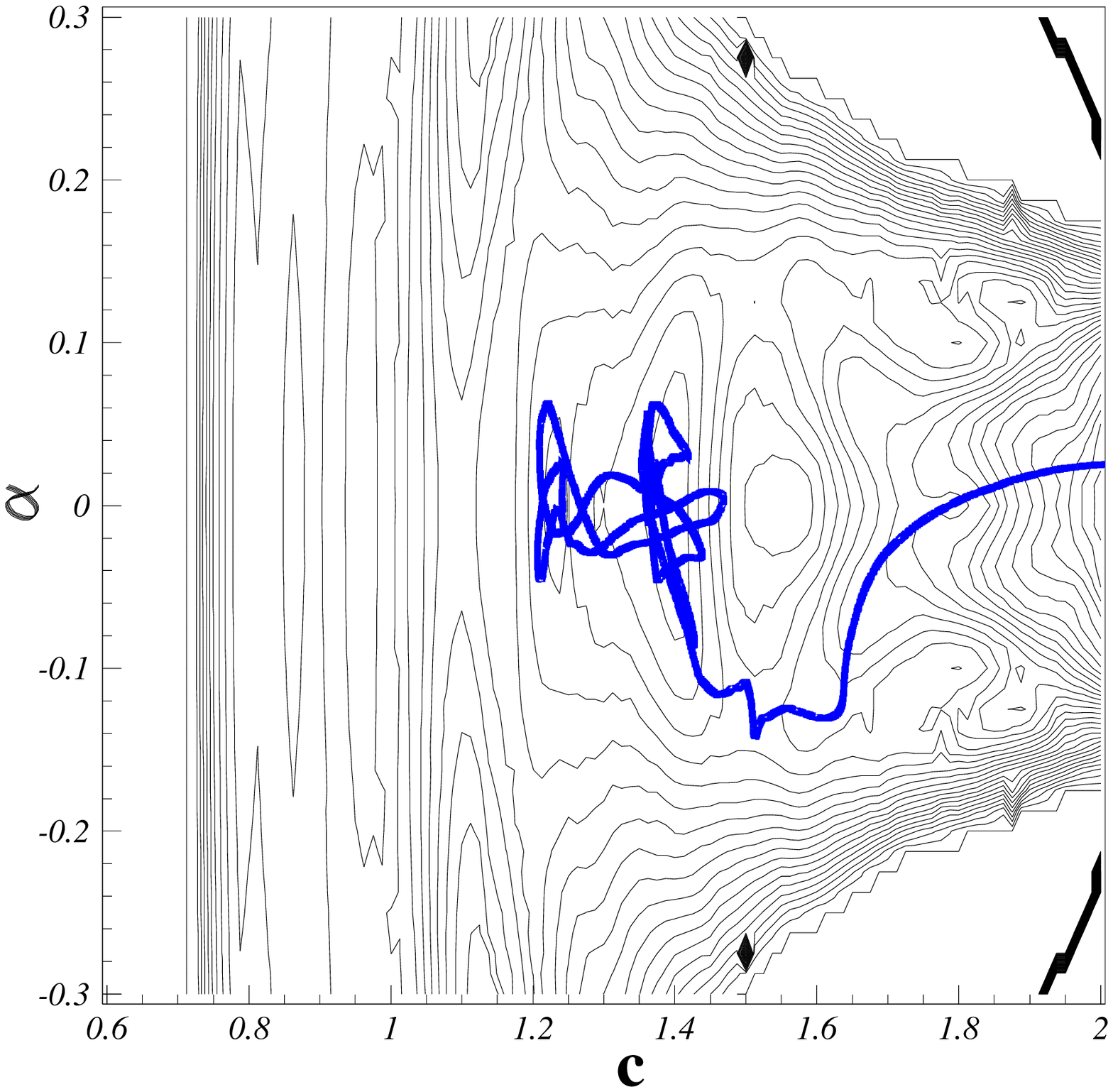,width=11.0cm,height=6.50cm}
\end{center}
\vspace{-0.5cm}
FIG.\ 9. Same as Fig.\ 6 with a typical asymmetric fission trajectory. 
\vspace{0.3cm}

In the LDM landscape the symmetric fission barrier at $L = 60$ MeV is of the 
order of $0.7$ MeV. On the other hand when quantal effects are included, the 
asymmetric fragment partition related to $\alpha \approx \pm 0.035$ corresponds 
to a barrier of $0.1$ MeV. One notices that in spite of a lower barrier height 
when including microscopic corrections the 
corresponding fission time is larger than in the semi-classical picture. One 
understands this result if we compare the typical symmetric trajectory drawn 
on Fig.\ 5 to the typical asymmetric trajectory drawn on Fig.\ 9. In the 
former case the system cannot overcome the rather high {\it mountain top} at 
$c \approx 1.55$ for $\alpha \in [-0.06 \,,\, 0.06]$ and consequently has to 
bypass it along $\alpha \approx \pm 0.1$ before reaching the asymmetric valley 
for $\alpha \approx \pm 0.035$. One could say that the path is {\it longer}. 
Let us note that a small energy difference (of the order of a 
few hundreds keV) bet\-ween  valleys can lead to very different fragment 
distributions what suggests the strong dependence of the dynamics on the 
{\it details} of the energy landscape. This drastic sensitivity to the
structure of the landscape requires to be careful when one performs energy 
calculations in the deformation space.
\subsection{Dynamics including light-particle evaporation}
Even if we know that particle evaporation is strongly reduced at low energy 
we have to admit that we do not have yet a complete description 
evaluating the emission widths $\Gamma_{\nu}$ at low temperature since one can 
seriously question the validity of the Weisskopf's theory
at such energies and since our development of the more microscopic 
phase-space distribution function approach is not complete. Nevertheless 
in order to investigate particle eva\-po\-ration, we consider in this 
section the compound nucleus  $^{227}$Pa at a higher total excitation energy 
of $56$ MeV for which we believe that the Weisskopf's 
approach should be approximately valid.
                                                                   \\[ -3.0ex]
\vspace{0.3cm}
\begin{center}
\begin{tabular}{||p{3.5cm}||*{2}{c|}|}
\hline
   & $\;\;\; V^{LDM} \;\;\;$ & $\;\; V^{LDM} + \delta E \;\;$ \\
\hline
\hline
$\;\; \sigma_{fis} / \sigma_{tot}$ ($\%$) & $99.8$ & $98.5$ \\
\hline
$\;\; {\bar t}_{fis} (\times 10^{-17} sec)$  & $2.335$ & $3.275$ \\ 
\hline
$\;\; {M}_{n}$ & $1.806$ & $2.153$ \\ 
\hline
$\;\; {M}_{p}$ & $0.010$ & $0.006$ \\ 
\hline
$\;\; {M}_{\alpha}$ & $0.017$ & $0.011$ \\ 
\hline
\end{tabular}
\end{center}
\vspace{0.1cm}
Table 1~: Influence of quantal corrections on fission probability, average 
fission time and light-particle multiplicities obtained for 
the system $^{227}$Pa ($E^*_{tot} \!=\! 56$ MeV, $L \!=\! 60 \hbar$).
                                                                   \\[ -1.0ex]

In Table 1 we compare the fission cross section, average fission time and 
light-particle multiplicities obtained  for the pure LDM description to the 
ones related to the potential energy surface including shell and pai\-ring 
corrections. As in the case without particle evaporation, one observes an 
increase of the fission 
time when quantal effects are taken into account.  Whereas the neutron 
pre-scission multiplicity is larger in the calculations with  
microscopic corrections, charged particle multiplicities are smaller. 
With Fig.\ 3 we have seen that neutrons can be emitted whatever the nuclear elongation, i.e. 
all along the fission path, and that their emission probability increases with increasing deformation. 
A longer fission time should therefore lead to a larger neutron multiplicity. Charged particles  
 are prefe\-rentially emitted at large 
deformations (see again Fig.\ 3). One has, however, to remember that when charged particle 
emission is favored a substantial amount of the 
available excitation energy of the emitting nucleus can already have been 
carried away through neutron emission. In addition one finds that the 
gradient of the potential energy for the asymmetric fission path including 
 quantal effects is larger between saddle and scission points  
than the one of the symmetric valley of the LDM landscape. 
This suggests that the corresponding time scale for the descent from saddle 
to scission is smaller in the case when shell and pairing effects are present 
what again favors a reduction of charged particle multiplicities. 
\subsection{Influence of excitation energy and angular momentum}
As shown in Fig.\ 10 an increase of the total excitation energy of the system 
from $26$ to $56$ MeV (which for a given angular momentum $L = 60 \hbar$ 
implies an increase of the thermal excitation energy) leads to a larger 
contribution to the symmetric fission mode. This result is obviously due to 
the vanishing of quantal effects when the nuclear temperature increases. 
However it can also be partly explained by a larger diffusion generated by 
the larger temperature (see Einstein relation). The corresponding larger 
oscillations thus allow the nucleus to {\it explore more easily} the energy 
landscape being able to overcome higher barriers and consequently to pass from 
one valley to another instead of being trapped preferentially in the deepest 
valley (which is asymmetric for the system presently considered).
In ref \ \cite{CS02} we also investigated the impact of the angular momentum 
on fission dynamics and obtained an relative increase of the symmetric fission 
cross section for increasing angular momentum due to a decrease of the 
fission barrier height.
\begin{center}
\epsfig{file=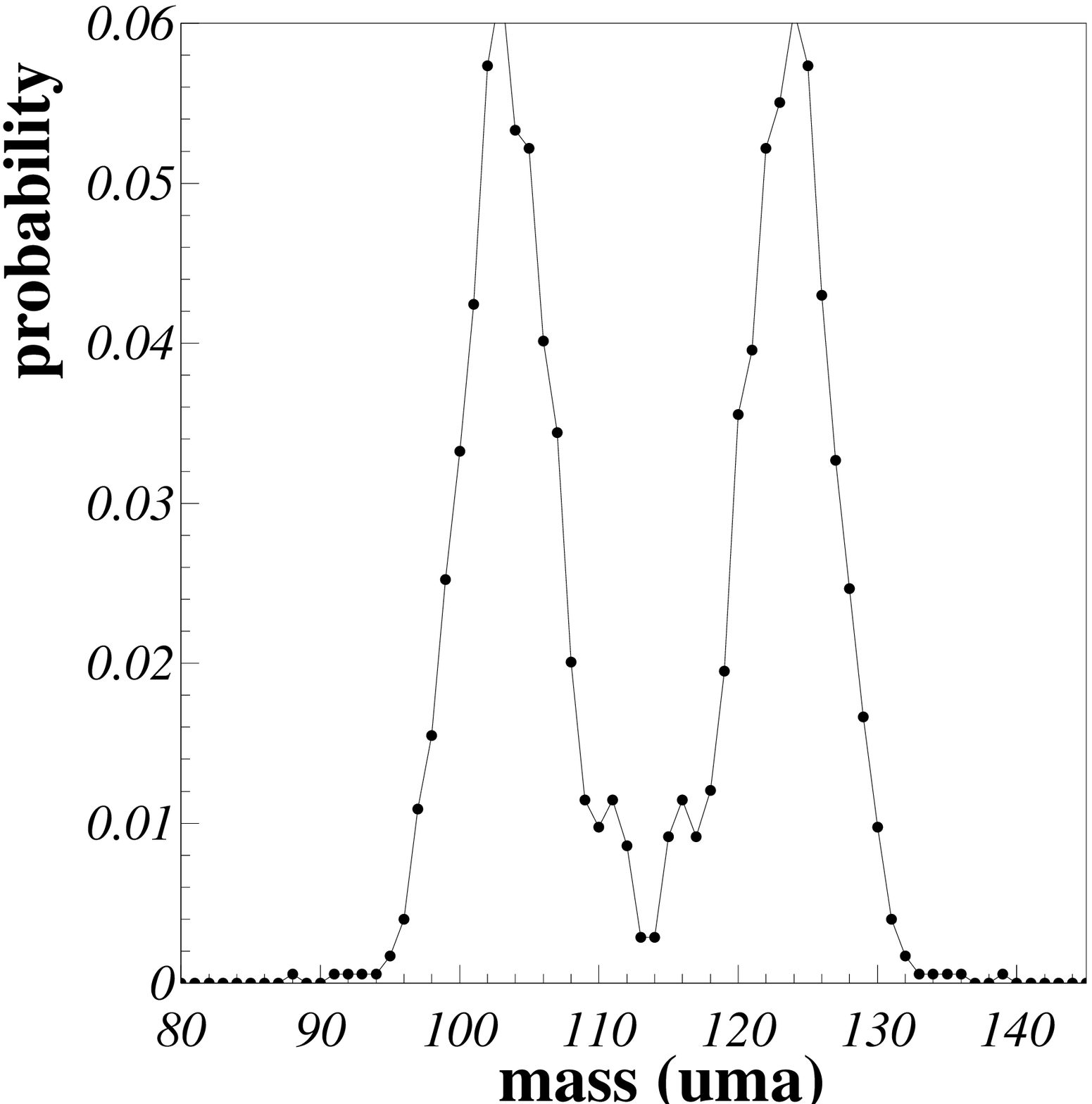,height=5.0cm,width=6.0cm}
\epsfig{file=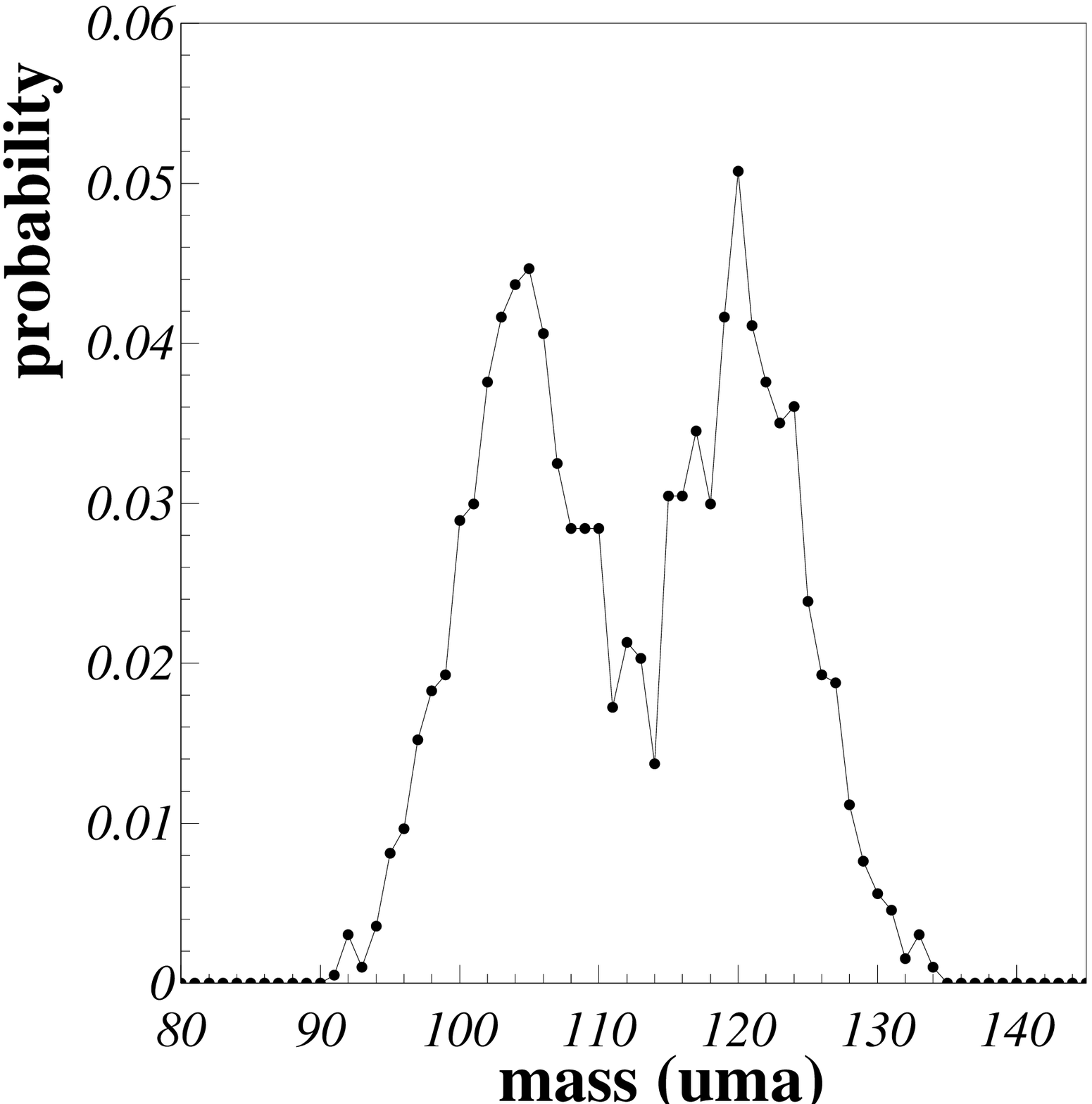,height=5.0cm,width=6.0cm}
\end{center}
\vspace{-5.5cm}
\hspace{1.9cm} {\large {\bf $E^*_{tot} = 26$ MeV}} \hspace{3.0cm} 
{\large {\bf $E^*_{tot} = 56$ MeV}} 
\vspace{4.8cm}
\\
FIG.\ 10. Fission-fragment mass distributions for two different 
values of the total excitation energy of the compound nucleus $^{227}$Pa 
at an angular momentum of $L = 60 \hbar$.
                                                                   \\[ -3.0ex]
\subsection{Evaluation of shell corrections close to the scission point}
In the framework of the Strutinsky method, shell correction calculations need 
to determine nuclear single particle levels which in our approach are the 
eigenvectors of a deformed Saxon Woods potential of standard parametrization 
\cite{BDJ72}.
In practice these states are obtained by an expansion in the basis of a 
deformed harmonic oscillator. This oscillator basis is an {\it one-center} 
basis which is probably not so well adapted if one is interested in describing 
shapes near the scission point. 
Indeed, for such strongly elongated and possibly {\it necked-in} surfaces, a 
{\it two-center} basis seems to be more adapted taking the structure of the 
nascent fragments better into account. We stress this technical detail in 
order to focus on the importance of a careful determination of shell effects 
at very large deformations. To illustrate this point we compare on Fig.\ 11 
the fragment mass distribution obtained when the dynamical calculation is 
artificially stopped at an elongation $c_{scis} = 1.8$ to the one obtained 
when this calculation is carried through up to the geometrical scission point 
$c_{scis} \!=\! c_{geo}$ where the splitting into two fragments takes place. 
The broad distribution related to $c_{scis} \!=\! 1.8$ can be easily 
understood with Fig.\ $6$ where the quite flat potential landscape in the 
$\alpha$ direction around $c \approx 1.8$ can give rise to a large variety of 
mass partitions. In spite of this, the final distribution at $c_{geo}$ is 
rather strongly asymmetric. Moreover the value $c \!=\! 1.8$ corresponds to a 
quite important elongation, i.e.\ an elongation for which one can already have 
a reasonable idea of the asymmetry of the nascent fission fragments 
\cite{NPPW79}. The present investigation points out the 
importance of quantal effects for $c \!>\! 1.8$ and with it the necessity of 
their accurate determination for these largest deformations. To avoid 
problems related to the choice of this one-center basis we perform the  
diagonalisation taking a very large number of basis states into account.
\vspace{-0.1cm}
\begin{center}
\epsfig{file=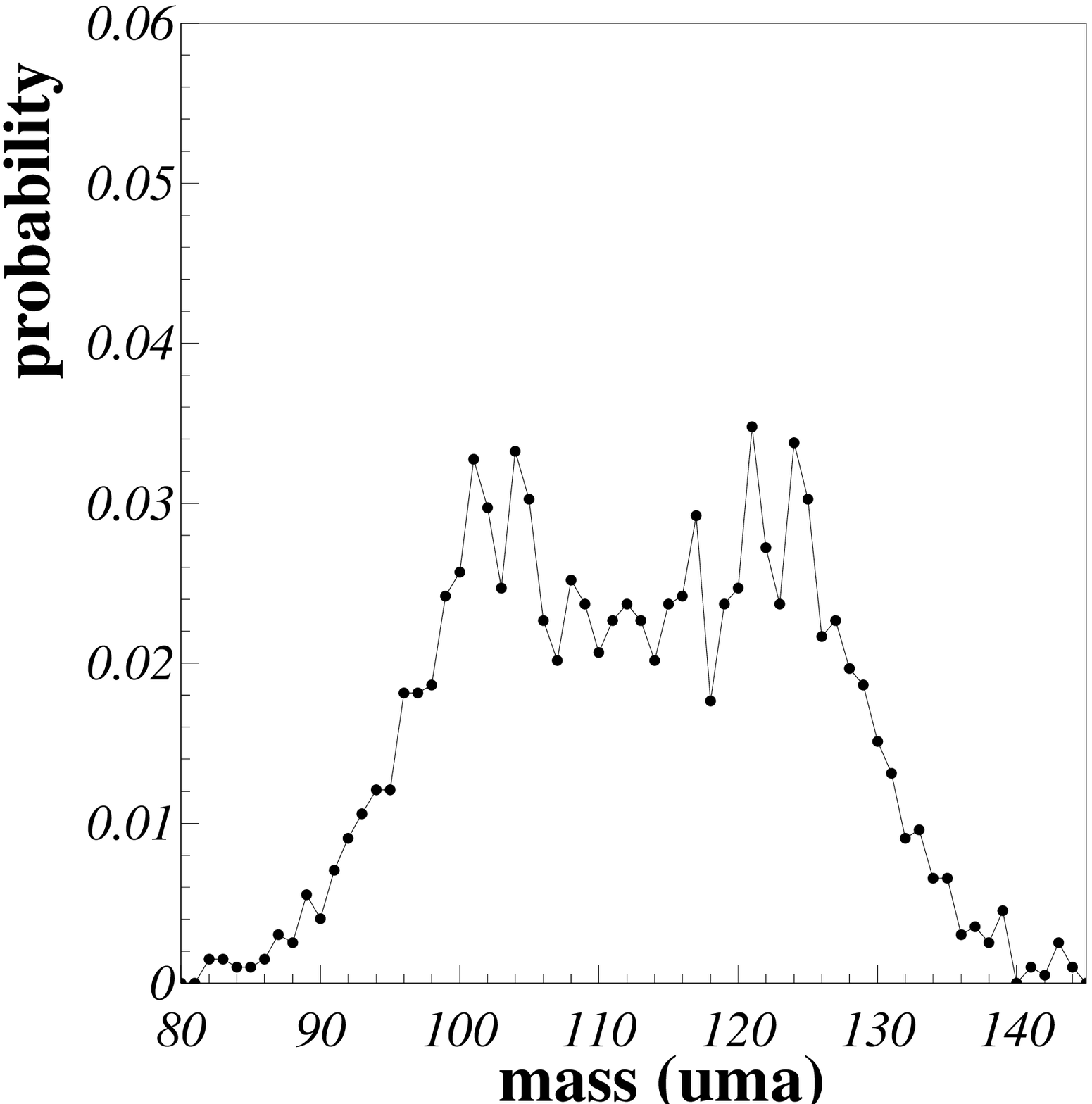,height=5.00cm,width=6.0cm}
\epsfig{file=mult-mas-qua-e56x.eps,height=5.00cm,width=6.0cm}
\end{center}
\vspace{-5.5cm}
\hspace{2.25cm} {\large {\bf $c_{scis} = 1.8$}} \hspace{4.05cm} 
{\large {\bf $c_{scis} = c_{geo}$}} 
                                                                   \\[ 31.0ex]
FIG.\ 11. Fission-fragment mass distributions obtained for the system $^{227}$Pa  
($E^*_{tot} = 56$ MeV, $L = 60 \hbar$) for $c_{scis} \!=\! 1.8$ and 
$c_{scis} \!=\! c_{geo}$ 
(see text).
                                                                   \\[ -3.0ex]
\subsection{Temperature dependence of transport coefficients}

As mentioned in section 2.2.\ we probably overestimate friction at low 
temperature. As demonstrated on Fig.\ 12 the 
reduction of friction by a factor of two ($0.5 \; w \& w$) results in a striking 
difference as compared to the full wall-and-window friction ($w \& w$). A larger friction 
causes a decrease of the kinetic energy of the system which
 is therefore  more sensitive to the {\it fine structure} of the landscape and consequently
 is more easily trapped in the deepest 
valleys. 
A smaller friction, on the contrary, allows the system, with larger kinetic energy, 
to move more 
freely through the landscape, to overcome more easily eventual barriers, 
resulting in a broader  
distribution. Reducing friction by a constant factor is obviously an extremely crude 
approximation to a real temperature dependent viscosity. 
We use this picture here simply to investigate the influence of  
friction on fragment distributions and light-particle multiplicities.
 
The procedure used in order to simulate in an approximate way the va\-nishing 
of quantal effects with temperature (see section 4.1.2) is still nowadays 
subject of controversies, in particular what pairing is concerned 
\cite{I78} - \cite{EgRo00}. Our investigations dealing with this point (for 
details see ref \ \cite{CS02}) showed that the T-dependence of shell and 
pairing corrections cannot be neglected, even if our system is already in the 
beginning of its decay at quite low excitation energy, which can still 
decrease along the fission path (namely due to particle evaporation).
\vspace{0.0cm}
\begin{center}
\epsfig{file=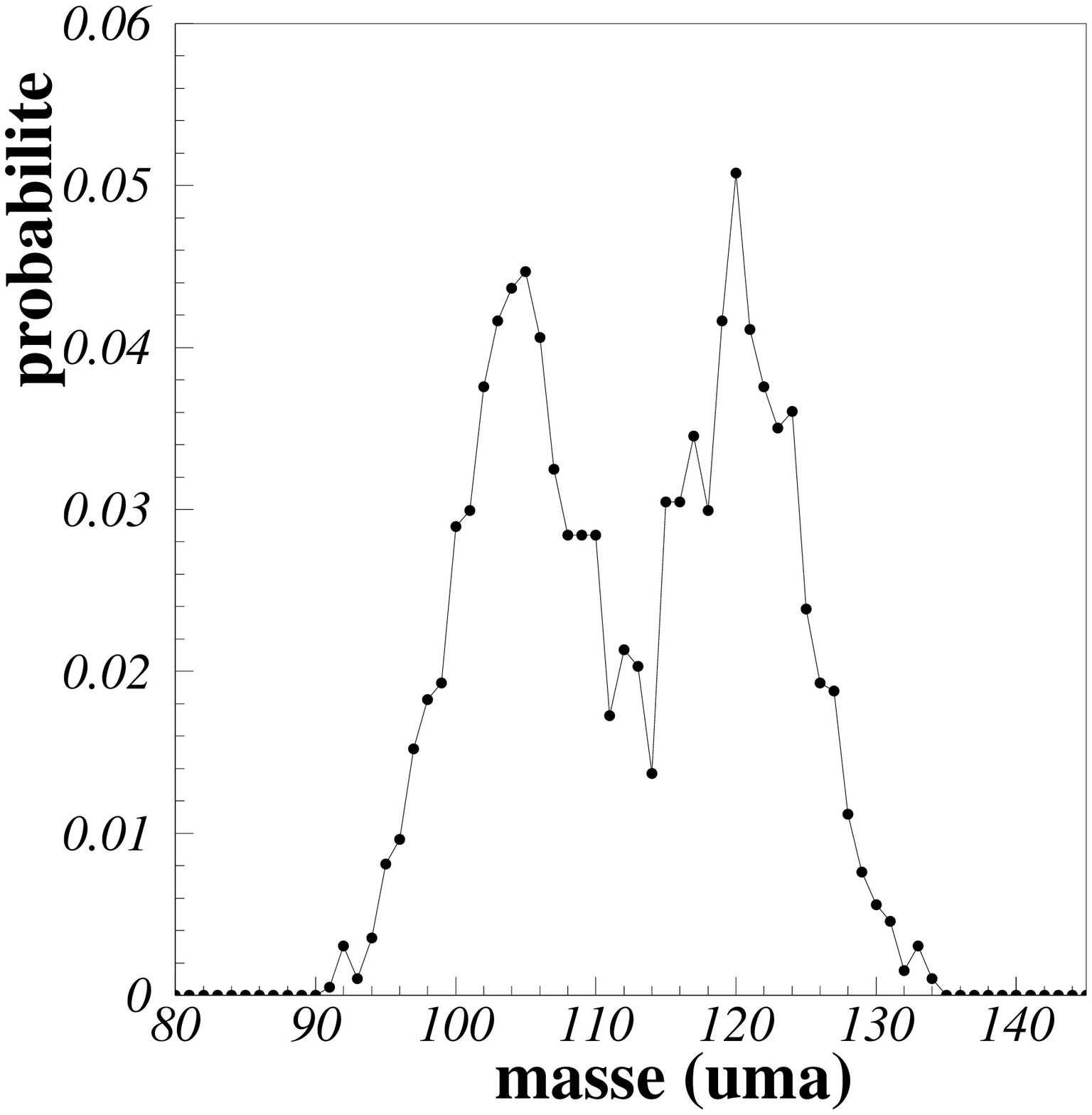,height=4.90cm,width=6.0cm}
\epsfig{file=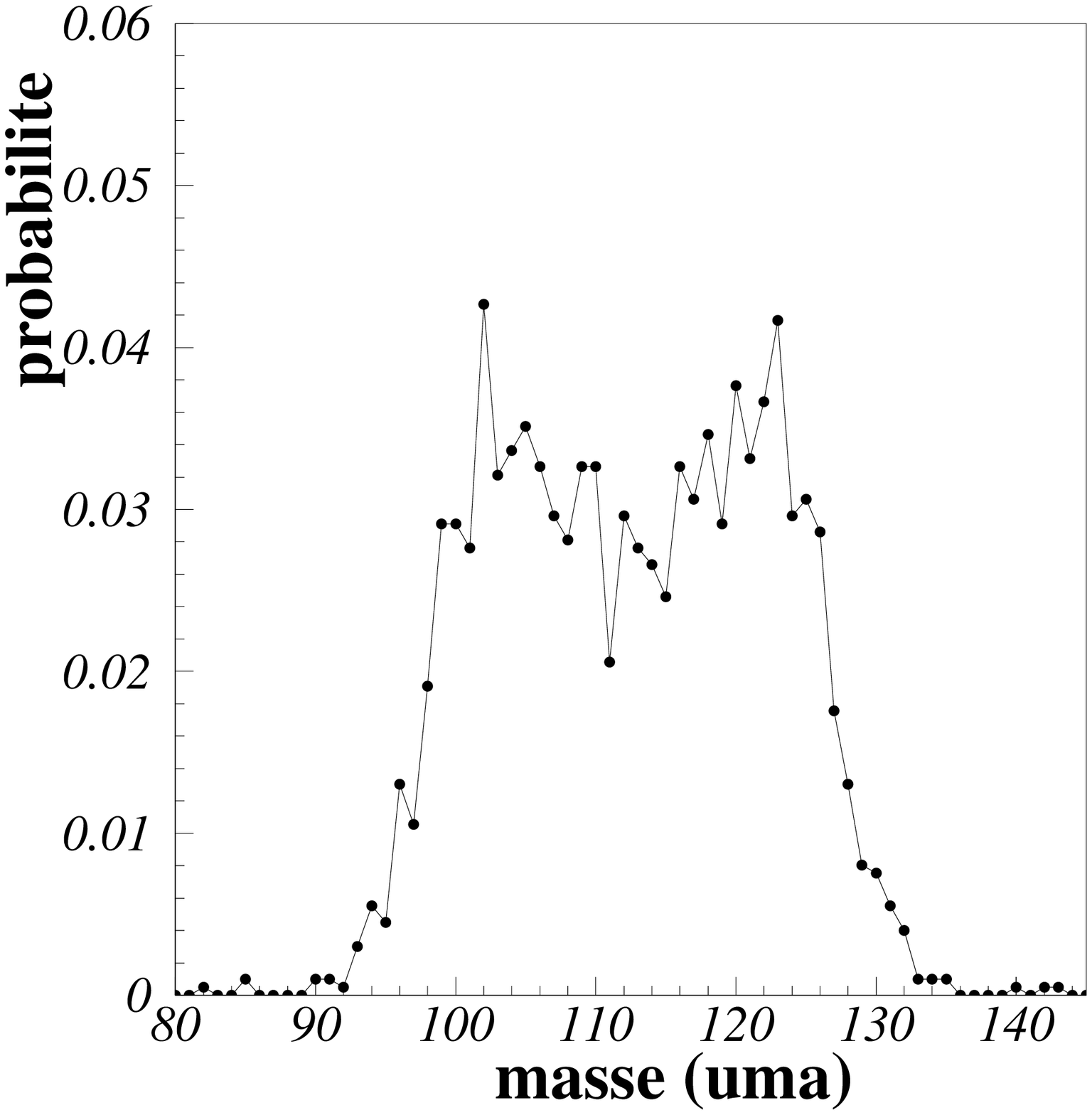,height=4.90cm,width=6.0cm}
\end{center}
\vspace{-5.35cm}
\hspace{2.7cm} {\large {\bf $w \& w$}} \hspace{4.6cm} {\large {\bf $0.5 \; w \& w$}} 
                                                                   \\[ 28.0ex]
FIG.\ 12. Fission-fragment mass distributions obtained for the system 
$^{227}$Pa ($E^*_{tot} \!=\! 56$ MeV, $L \!=\! 60 \hbar$) with the full 
($w \& w$) and a reduced ($0.5 \; w \& w$) friction (see text).
                                                                   \\[ -3.0ex]
%\newpage
%
%\vspace{0.1cm}
%\begin{center}
%\begin{tabular}{||p{3.60cm}||*{2}{c|}|}
%\hline
%  & $\;\;$ $w \& w$ $\;\;$ & $\;\;$ $0.5 \; w \& w$ $\;\;$ \\
%\hline
%\hline
% $\;\;\; \sigma_{fis}$ / $\sigma_{tot}$ ($\%$) & 98.5 & 99.6 \\
%\hline
%$\;\;\; {\bar t}_{fission} (\times 10^{-17})$ s & 3.275 & 2.194 \\ 
%\hline
% $\;\;\;$ M$_{n}$ & 2.153 & 1.767 \\ 
%\hline
%\end{tabular}
%\end{center}
%\vspace{0.05cm}
%
%
%Table 2~: Influence of the strength of friction on some fission characteristics.
%                                                                    \\[ -1.0ex]
%								   
%In Table 2 we display some fission feature for the full and the reduced 
%frictions.  The increase of the prescission neutron multiplicity is explained by the increase of the fission
% time in the case of the larger viscosity.
%
\section{Confrontation with experimental data}

As the agreement theory-experiment at high excitation energy is quite 
promising \cite{PNS00}, we would like to compare in the present section  
our  predictions to the available experimental data concerning the fission
process of the nucleus $^{227}$Pa synthesized at a total excitation energy of 
$E^*_{tot} \!=\! 26$ MeV \cite{CS02}. In the calculations we should obviously 
take particle evaporation into account. Since we do not have for the moment 
a complete reliable evaporation theory at our disposal at low temperature 
we first performed dynamical calculations at higher energy for which we 
believe that the Weisskopf's approach is about reasonable. This study showed 
us that the influen\-ce of particle evaporation on the fission fragment mass 
distribution can be neglected \cite{CS02}. As the probability of emitting 
particles decreases with excitation ener\-gy \cite{PBR96}, we also expect a 
really small impact of evaporation on the mass distribution at $26$ MeV. 
Consequently we compare in Fig.\ 13 mass distributions obtained for 
$E^*_{tot} \!=\! 26\,$MeV without taking evaporation into account with the 
experimental mass distribution.
We have considered in the theo\-retical calculations three different 
frictions~: $25 \%$ of the wall and window value, $20 \%$ and $15 \%$.
								    
The experimental analysis has exhibited a multi-modal fission-fragment mass 
distribution \cite{CS02} composed of three modes~: the symmetric one and two 
asymmetric modes centered around mass $A \!=\! 132$ corresponding to the 
double magic $^{132}$Sn nucleus and around mass $A \!=\! 140$ related to the 
deformed $^{140}$Ba nucleus, explained \cite{BG90,IVP00} by the closure of the 
deformed neutron shell $N \!=\! 84$. The comparison with our predictions shows 
that in the case of a friction corresponding to $15 \%$ of the wall-and-window 
value the model reproduces quite well the symmetric fission mode. We would 
like to mention here that microscopic 
calculations performed by Hofmann and Ivanyuk \cite{HIR01} indicate that 
such a reduced viscosity is about what is to be expected at such low 
excitation energy. However our calculation gives only rise to the asymmetric 
$A \!=\! 132$ channel, the $A = 140$ mode being absent.
                                                                   \\[  2.0ex]
${}$ \hspace{0.5cm} $\;\; 0.25 \; w \& w 
\;\;\;\;\;\;\;\;\;\;\;\;\;\;\;\;\;\;\;\;\;\;\;\;\;\; 0.2 \; w \& w
\;\;\;\;\;\;\;\;\;\;\;\;\;\;\;\;\;\;\;\;\;\;\;\;\;\; 0.15 \; w \& w$  \\[-6.0ex]
%	{\tiny .} \hspace{1.2cm}	
\begin{center}
\hspace*{-0.6cm}\epsfig{file=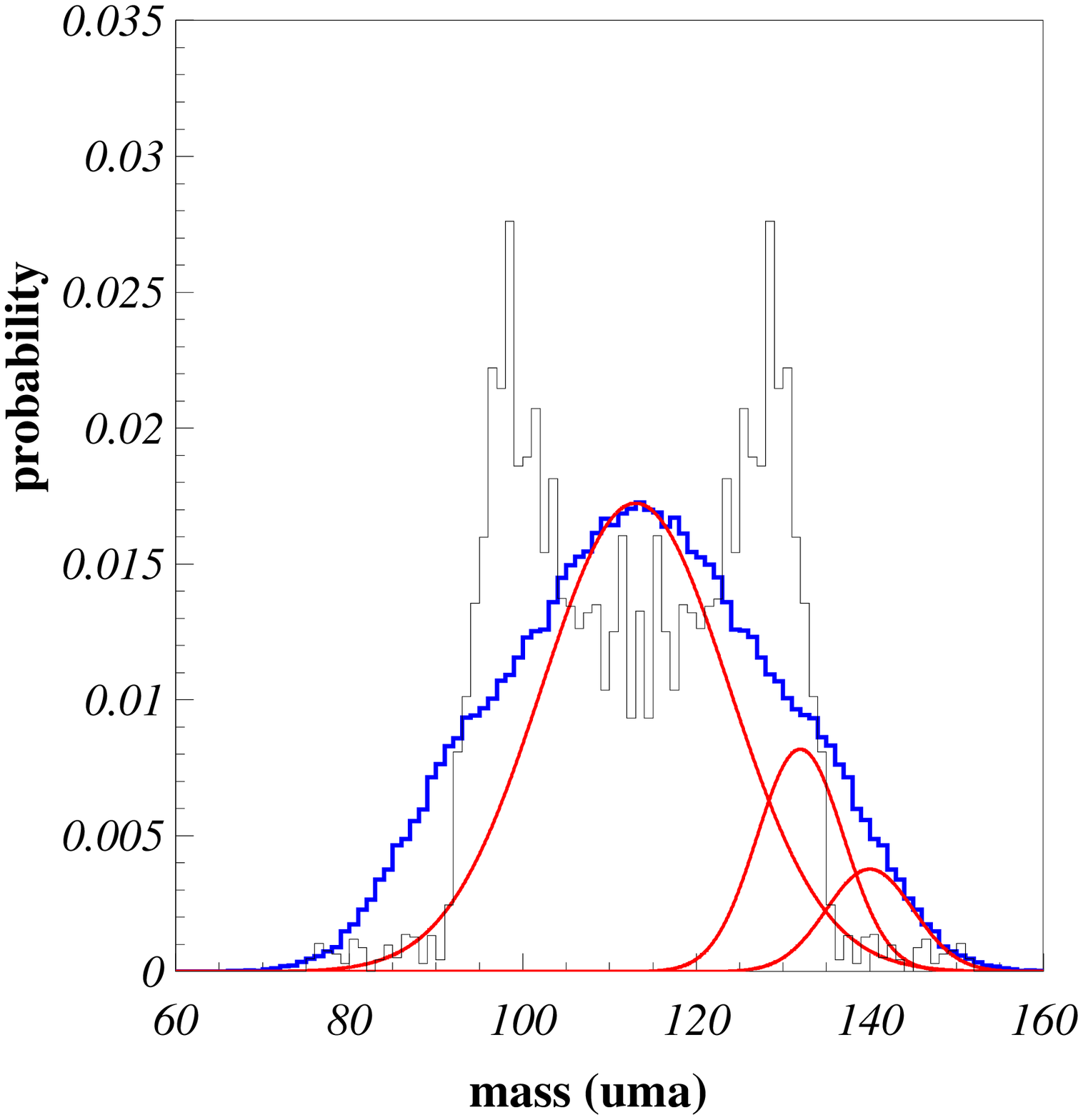,width=4.5cm,height=5.cm}
\vspace*{-5.0cm}
\hspace*{-0.3cm}\epsfig{file=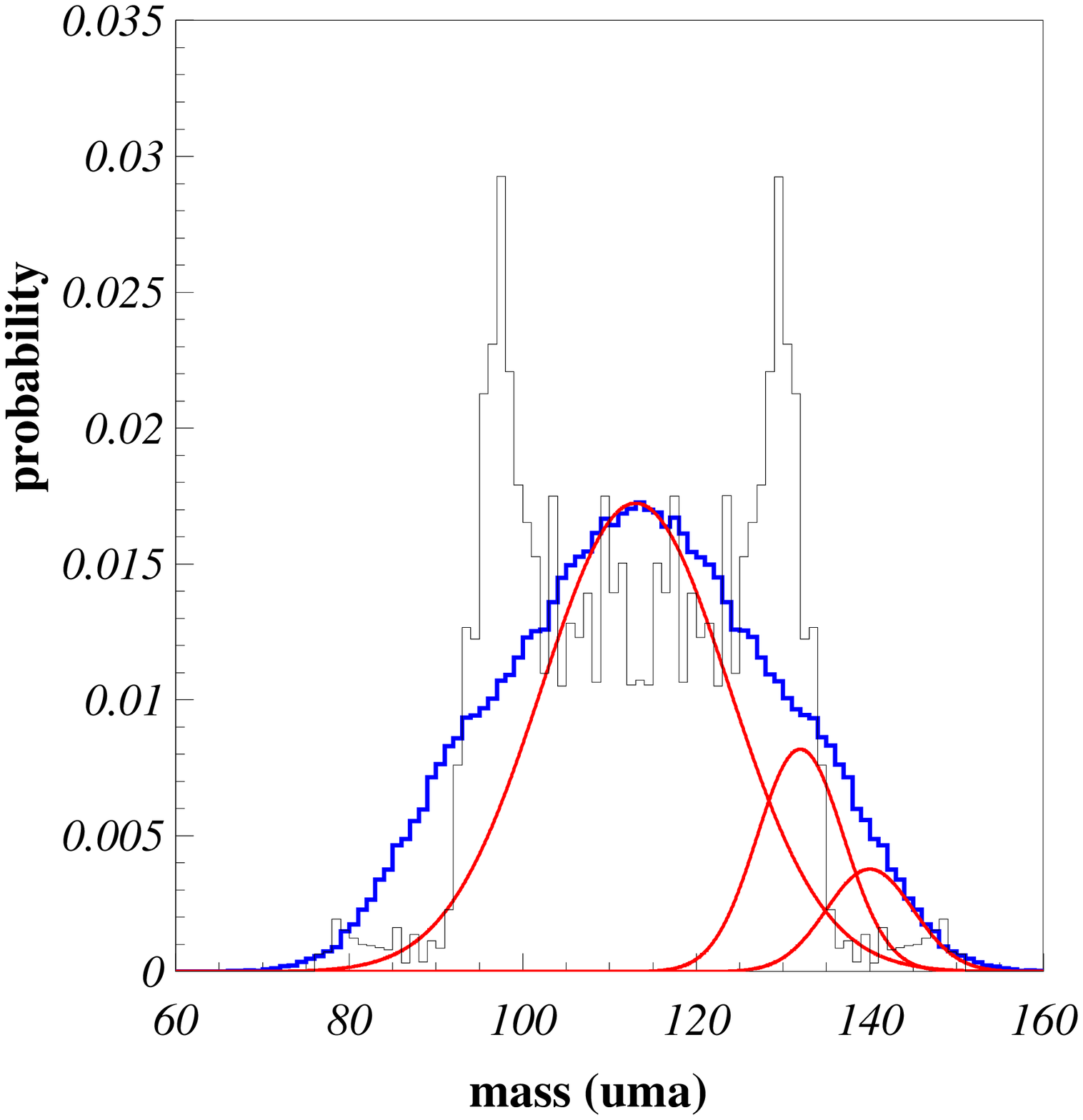,width=4.5cm,height=5.cm}
\vspace*{-5.0cm}
\hspace*{-0.3cm}\epsfig{file=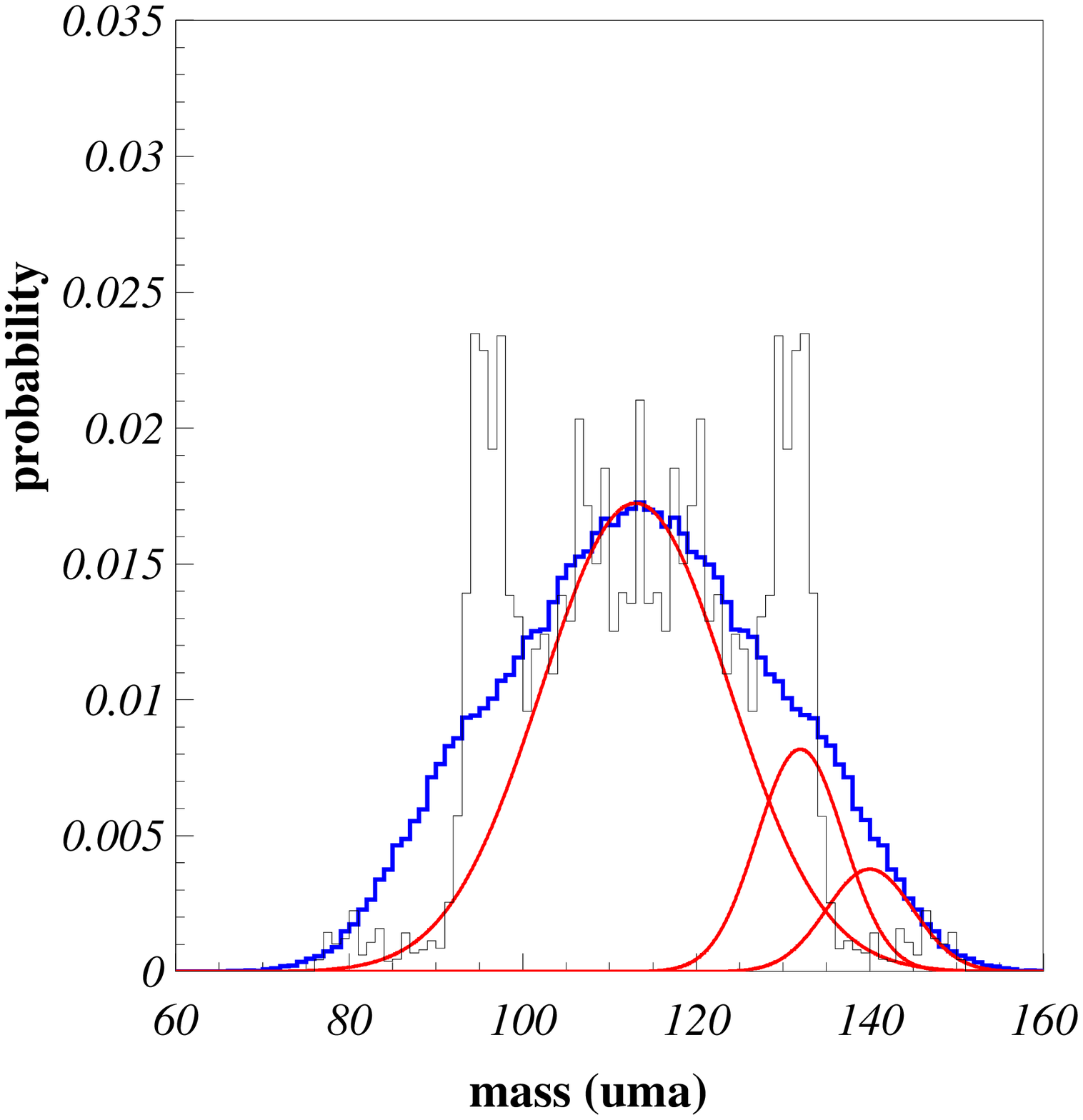,width=4.5cm,height=5.cm}
%
%\hspace*{-0.8cm}\epsfig{file=pour_fig/compar_0.2ww_sans_cgeombis.eps,width=4.5cm,height=5.cm}
%\vspace*{-5.0cm}
%\hspace*{-0.3cm}\epsfig{file=pour_fig/compar_0.2ww_sans_cgeombis.eps,width=4.5cm,height=5.cm}
%\vspace*{-5.0cm}
%\hspace*{-0.3cm}\epsfig{file=pour_fig/compar_0.15ww_sans_cgeombis.eps,width=4.5cm,height=5.cm}
\end{center}
\vspace{9.8cm}
FIG.\ 13. Experimental (solid line) and theoretical (histograms) fission 
fragment mass distributions for the system $^{227}$Pa ($E^*_{tot} = 26$ MeV) 
for different values of the friction.
                                                                   \\[ -0.50ex]
 
In order to understand the disagreement between our model and the experimental 
data for asymmetric fission we have to remember that we have chosen to 
describe nuclear shapes in the 2-dimensional deformation space ($c, \alpha$) 
imposing $h \!=\! 0$. Taking $h$ different from zero will allow us to consider  
a larger variety of nuclear configurations. We thus believe that with the 
res\-tricted 2-dimensional parametrization we are not able to give a description of the 
deformed shape of $^{140}$Ba but that when taking $h \!\neq\! 0$ into account 
we will describe that shape and the corresponding asymmetric fission valley so 
that a part of the trajectories which, for $h \!=\! 0$, end up in the 
$A \!=\! 132$ channel will reach, in the case of $h \neq 0$, the previously 
missing $A \!=\! 140$ valley. 
The contribution to the fission mode $A \!=\! 132$ will then decrease while the 
one of the $A = 140$ channel will increase, thus reaching a better agreement 
between theory and experiment when we will have extended the present 
2-dimensional treatment to a 3-dimensional one. Investigations along 
this direction are under way.
                                                                   \\[ -3.0ex]

%
%%%%%%%%%%%%%%%%%%%%%%%%%%%%%%%%%%%%%%%%%%%%%%%%%%%%%%%%%%%%%%%%%%%%%%%%%%%%%%
%%%%%%%%%%%%%%%%%%%%%%%%%%%%%%%%%%%%%%%%%%%%%%%%%%%%%%%%%%%%%%%%%%%%%%%%%%%%%%
%%%%%%%%%%%%%%%%%%%%%%%%%%%%%%%%%%%%%%%%%%%%%%%%%%%%%%%%%%%%%%%%%%%%%%%%%%%%%%
%
\section{Discussion and conclusions}
With the purpose to study multi-modal fission, we have developed a model 
describing the dynamics of the fission process by the resolution of a 
2-dimensional Langevin equation coupled to the Master equations go\-verning 
particle emission. Starting from a more or less classical description proven 
as rather successful for describing symmetric fission at high excitation 
energy, we extended our theory to multi-modal fission by increasing the 
dimensionality of the deformation space in which the Langevin equation is 
solved in order to be able to deal with asymmetric shapes and by including 
quantal effects (shell and pairing corrections) in the potential-energy 
calculations. Our investigations show the strong sensitivity of the dynamics 
on the structure of the potential-energy landscape what implies the necessity 
for a careful description of the later, in particular in the determination of 
shell and pairing corrections at large deformation.

Comparing theoretical and experimental fission-fragment mass distributions 
one observes a rather promising agreement which, as we believe, could still 
be considerably improved if the 2-dimensional treatment is extended to
a 3-dimensional one. We also point out the importance of taking into account 
the temperature dependence of nuclear friction which as we have seen should 
be significantly reduced at low energy. Another crucial aspect of the problem 
lies in the necessity of a reliable evaporation theory at low excitation 
energy.
 
Up to now the general analysis was that pre-scission light-particle 
multiplicities were the quantities to investigate \cite{Fro98,PNS00} for a better 
understanding of fission dynamics. Our present study shows, on the contrary, 
that at low excitation energies where the number of emitted particles is small  
and, in the frequent case where the competition between symmetric and 
asymme\-tric channels exhibits multi-modal fission, the fragment mass distribution 
is probably more relevant, in particular for investigating transport
coefficients  like nuclear friction. 
                                                                   \\[  3.0ex]
{\bf Acknowledgments}
                                                                   \\[ -1.0ex]
 
Two of us (C.S. and J.B.) are very grateful for the hospitality extended 
to them on many occasions by the Theory Department of the 
Marie-Curie-Sk\l odowska University in Lublin. A.S. and K.P. are in turn 
very thankful for many fruitful visits they were able to do to the 
Strasbourg Nuclear Research Center IReS and its Nuclear Theory group. 
K.P. especially acknow\-ledges a PAST position granted to him by the 
University Louis Pasteur and the French Ministry of National Education and 
Research. This work was partially sponsored by the Polish Committee of 
Scientific Research KBN No.~2P~03B~115~19 and the POLONIUM fellowship
No.~007/IN2P3/4788/2002.
 
%%%%%%%%%%%%%%%%%%%%%%%%%%%%%%%%%%%%%%%%%%%%%%%%%%%%%%%%%%%%%%%%%%%%%%%%%%%%%
%%%%%%%%%%%%%%%%%%%%%%%%%%%%%%%%%%%%%%%%%%%%%%%%%%%%%%%%%%%%%%%%%%%%%%%%%%%%%
%%%%%%%%%%%%%%%%%%%%%%%%%%%%%%%%%%%%%%%%%%%%%%%%%%%%%%%%%%%%%%%%%%%%%%%%%%%%%
                                                                                %

%\newpage


\begin{thebibliography}{99}
\bibitem{GPW79} P. Grang\'e, H.C. Pauli and H.A. Weidenm\"uller, 
                Phys.\ Lett.\ {\bf B88} (1979) 9; \\
                Z.\ Phys.\ {\bf A296} (1980) 107.
\bibitem{SDP91} E. Strumberger, W. Dietrich and K. Pomorski, 
                Nucl.\ Phys.\ {\bf A529} (1991) 522.
\bibitem{Prz94} W. Przystupa, K. Pomorski, Nucl. Phys. {\bf A572} (1994) 153.
\bibitem{AAR96} Y. Abe, S. Ayik, P.-G. Reinhard and E. Suraud, Phys. Rep. 
                {\bf 275} (1996) 49.
\bibitem{PBR96} K. Pomorski, J. Bartel, J. Richert and K. Dietrich,
                Nucl. Phys. {\bf A605} (1996) 87.
\bibitem{Fro98} P. Fr\"obrich, I.I. Gontchar, Phys. Rep. {\bf 292} (1998) 131.
\bibitem{PoStru90} K. Pomorski, E. Strumberger, Ann.Univ.MCS, Poland, 
 sec.AAA XLV (1990) 113.
\bibitem{TKS80} S. Trentalange, S.E. Koonin, A. Sierk, Phys.\ Rev.\ 
                {\bf C22} (1980) 1159.
%\bibitem{BDJ72} M. Brack, et al. Rev.\ Mod.\ Phys.\ {\bf 44} (1972) 320.
%\bibitem{Pau73} H.-C.\ Pauli, Phys.\ Rep.\ {\bf 7} (1973) 44.
\bibitem{BDJ72} M. Brack, J. Damgaard, A.S. Jensen, H.C. Pauli, 
             V.M. Strutinsky, C.Y.Wong, Rev.\ Mod.\ Phys.\ {\bf 44} (1972) 320.
\bibitem{Da76}  K.T.R. Davies, A.J. Sierk, J.R. Nix, Phys.\ Rev.\ 
                {\bf 13C} (1976) 2385.
\bibitem{BRS77} J. Blocki, J. Randrup, W. J. Swiatecki and C.W. Tsang, 
                Ann. Phys. {\bf 105} (1977) 427.
\bibitem{Fe87}  H. Feldmeier, Rep. Prog. Phys. {\bf 50} (1987) 915.
\bibitem{BMR96} J. Bartel, K. Mahboub, J. Richert and K. Pomorski, 
                Z.\ Phys.\ {\bf A354} (1996) 59.
\bibitem{PH91} K. Pomorski, H. Hofmann, Phys. Lett. {\bf  B263} (1991) 164.
\bibitem{HoIv99} H. Hofmann, F.A. Ivanyuk, Phys. Rev. Lett. {\bf  82} (1999) 4603.
\bibitem{S67}   V.M. Strutinsky, Nucl. Phys. {\bf A95} (1967) 420; 
                {\bf A122} (1968) 1.
\bibitem{BCS57} J.\ Bardeen, L.N.\ Cooper and J.R.\ Schrieffer, 
                Phys.\ Rev.\ {\bf 108} (1957) 1175.
\bibitem{NTS69}  S.G. Nilsson, C.F. Tsang, A. Sobiczewski, Z. Szymanski, S. Wycech, C. Gustafson, 
                I.-L. Lamm, P. Moller, B. Nilsson, Nucl. Phys. {\bf  A131} (1969) 1.
\bibitem{PNS00} K.~Pomorski, B.~Nerlo-Pomorska, A.~Surowiec, M.~Kowal,
                J.~Bartel, K.~Dietrich, J.~Richert, C. Schmitt, B.~Benoit, 
                E.~de~Goes~Brennand, L.~Donadille, C.~Badimon,
                Nucl. Phys. {\bf A679} (2000) 25.
\bibitem{CSLi00}  C. Schmitt, A. Surowiec, J. Bartel, K. Pomorski, Proccedings of the Int. 
Conf. Nuclear Physics at Border Lines, Lipari, Italy, 
May 21-24, 2001, Edited by G.Fazio, G.Giardina, F.Hanappe, G.Immè and N.Rowley, 
World Scientific.
\bibitem{CS02} C. Schmitt, PhD Thesis, Universit\'e Louis Pasteur 
               Strasbourg ($2002$), IReS $02-04$.
\bibitem{Wei37} V. Weisskopf, Phys. Rev. {\bf 52} (1937) 295.
\bibitem{DPR95} K. Dietrich, K. Pomorski and J. Richert, 
                Z.\ Phys.\ {\bf A351} (1995) 397.
\bibitem{SPS01}  A. Surowiec, K. Pomorski, C. Schmitt, Acta Physica Polonica {\bf B33} (2002).
in Nuclear Fission, Contributed Papers, W. Berlin ($1989$), 63.
%\bibitem{Yam98} S. Yamaji, H. Hofmann, R. Samhammer, Nucl. Phys. {\bf A475}
%                (1998) 487.
\bibitem{I78}  A.V. Ignatyuk, Phys. Lett. {\bf  B76} (1978) 543.
\bibitem{I80}  A.V. Ignatyuk, Nucl. Phys. {\bf  A346} (1980) 191.
\bibitem{IST75} A.V. Ignatyuk, G.N. Smirenkin, A.S. Tishin, Yad. Fiz. {\bf  21} (1975) 485.
\bibitem{EgRo00} J.L. Egido, L.M. Robledo, V. Martin, Phys. Rev. Lett. \ {\bf 85} (2000) 054308.
\bibitem{MN70} P. Moller, S.F. Nilsson, Phys. Lett. \ {\bf B31} (1970) 283.
\bibitem{MG74} J.M. Maruhn, W. Greiner, Phys. Rev. Lett. \ {\bf 32} (1974) 548.
\bibitem{NPPW79} B. Nerlo-Pomorska, K. Pomorski, E. Werner, International Conference Fifty Years Research
\bibitem{BG90} U. Brosa, S. Grossman, A. Muller,  Physics Reports, {\bf 197} 
               (1990) 167.
\bibitem{IVP00} I.V. Pokrovsky et al., Phys. Rev. {\bf C62} (2000) 014615.
\bibitem{HIR01} H. Hofmann, F.A. Ivanyuk, C. Rummel, S. Yanji,  
                Phys.\ Rev.\ {\bf C64} (2001) 054316.
\end{thebibliography}
\end{document}